%% file: main.tex
\documentclass[traditabstract]{aa} 
\usepackage{graphics,color}
\usepackage{aas_macros}
\usepackage{afterpage}
\usepackage{subfigure}
\usepackage{enumerate}
\usepackage{epsfig}
\usepackage{txfonts}
\usepackage{float}
\usepackage{natbib}
\usepackage{url}
\usepackage{bm}
\usepackage[latin1]{inputenc}

\begin{document}
	\title{Stochastic orbital migration of small bodies in Saturn's rings}
	\titlerunning{Stochastic orbital migration of moonlets in Saturn's rings}

\author{Hanno Rein \and John C. B. Papaloizou }
\authorrunning{H. Rein and J.C.B. Papaloizou}

\institute{	University of Cambridge, Department of Applied Mathematics and Theoretical Physics,\\
		Centre for Mathematical Sciences, Wilberforce Road, Cambridge CB3 0WA, UK\\
\email{hr260@cam.ac.uk}
}
\abstract{
Many small moonlets, creating propeller structures, have been found in Saturn's rings by the Cassini spacecraft.
We study the dynamical evolution of such 20-50m sized bodies which are embedded in Saturn's rings.
We estimate the importance of various interaction processes with the ring particles on the moonlet's eccentricity
and semi-major axis analytically.
For low ring surface densities, the main effects on the evolution of the eccentricity and the semi-major axis are
found to be due to collisions and the gravitational interaction with particles in the vicinity of the moonlet. 
For large surface densities, the gravitational interaction with self-gravitating wakes becomes important.

We also perform realistic three dimensional, collisional N-body simulations with up to a quarter of a million particles.
A new set of pseudo shear periodic boundary conditions is used which reduces the computational costs by an order of 
magnitude compared to previous studies. Our analytic estimates are confirmed to within a factor of two.

On short timescales the evolution is always dominated by stochastic effects caused by collisions 
and gravitational interaction with self-gravitating ring particles.
These result in a random walk of the moonlet's semi-major axis. 
The eccentricity of the moonlet quickly reaches an equilibrium value due to collisional damping.
The average change in semi-major axis of the moonlet after 100 orbital periods is 10-100m. 
This translates to an offset in the azimuthal direction of several hundred kilometres. 
We expect that such a shift is easily observable.
}
\keywords{Saturn -- planetary rings -- moons -- satellites -- celestial mechanics -- collisional physics -- N-body simulations}

\maketitle

\section{Introduction}
Small bodies, so called moonlets, which are embedded in Saturn's rings, can create propeller shaped structures due to their disturbance of the rings. These have been predicted both analytically and numerically \citep{Spahn2000,Sremcevic2002,Seiss2005}. 
Only recently, they have been observed by the Cassini spacecraft in both the A and B ring \citep{Tiscareno2006, Tiscareno2008}. 

These $20\,\mathrm{m}\,-100\,\mathrm{m}$ sized bodies can migrate within the rings, similar to proto-planets which migrate in a proto-stellar disc. 
Depending on the disc properties and the moonlet size, this can happen in either a smooth or in a stochastic (random walk) fashion. 
We refer to those migration regimes as type I and type IV, respectively, in analogy to the terminology in disc-planet interactions. 
\cite{Crida2010} showed that there is a laminar type I regime that might be important on very long timescales.
This migration is qualitatively different to the migration in a pressure supported gas disc.
However, the migration of moonlets in the A ring is generally dominated by type IV migration, at least on short timescales. 

In this paper, we study the type IV migration regime in detail.
The full mutual interaction of the ring particles with the moonlet and its consequent induced motion are considered both analytically and numerically.

We first review the basic equations governing the moonlet and the ring particles in a shearing box approximation in Sect. \ref{sec:analytic}.
Then, in Sect. \ref{dampe}, we estimate the eccentricity damping timescale due to ring particles colliding with the moonlet and ring particles on horseshoe orbits as well as the effect of particles on circulating orbits.
In Sect. \ref{sec:excitee} we estimate the excitation of the moonlet eccentricity caused by stochastic 
particle collisions and gravitational interactions with ring particles.
This enables us to derive an analytic estimate of the equilibrium eccentricity.

In Sect. \ref{sec:randa} we discuss and evaluate processes, such as collisions and gravitational interactions
with ring particles and self-gravitating clumps, that lead to a random walk in the semi-major axis of the moonlet. 

We describe our numerical code and the initial conditions used in Sect. \ref{sec:numerical}. 
We perform realistic three dimensional N-body simulations of the ring system and the moonlet, taking into account a moonlet 
with finite size, a size distribution of ring particles, self-gravity and collisions.
The results are presented in Sect. \ref{sec:results}. 
All analytic estimates are confirmed both in terms of qualitative trends
and quantitatively to within a factor of about two in all simulations that we performed. 
We also discuss the long term evolution of the longitude of the moonlet and its observability,
before summarising our results in Sec. \ref{sec:conclusions}.

\section{ Basic equations governing the moonlet and ring particles}\label{sec:analytic}
\begin{figure*}
\center
\resizebox{0.77\textwidth}{!}{
\input{trajectory.pstex_t}
}
	\caption{Trajectories of ring particles in a frame centred on the moonlet. Particles accumulate near the moonlet and fill it's Roche lobe. Particles on trajectories labelled a) are on circulating orbits. Particles on trajectories labelled b) collide with other particles in the moonlet's vicinity.
	Particles on trajectories labelled (c) collide with the moonlet directly.
	Particles on trajectories labelled (d) are on horseshoe orbits. Particles on trajectories
	labelled (e) leave the vicinity of the moonlet. \label{fig:trajectory}}
\end{figure*}
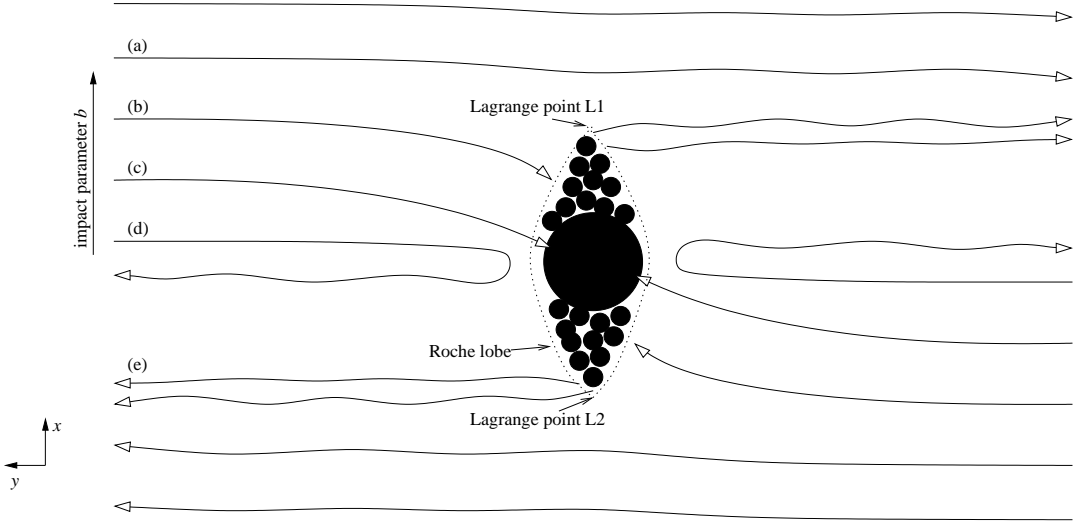

We adopt a local right handed Cartesian coordinate system with its origin being in circular Keplerian orbit with semi-major axis $a$ and 
rotating uniformly
with angular velocity $\Omega.$
This orbit coincides with that of the moonlet when it is assumed to be 
unperturbed by ring particles. The $x$ axis coincides with the line joining the 
central object of mass $M_p$ and the origin. The unit vector in the $x$ direction, $\mathbf{e}_x,$ points away
from the central object. The unit vector in the $y$ direction $ \mathbf{e}_y$ points in the direction
of rotation and the unit vector in the $z$ direction, $\mathbf{e}_z$ points in the vertical direction
being normal to the disc mid-plane. 

In general, we shall consider a ring particle of
mass $m_1$ interacting with the moonlet which has a much larger mass $m_2.$ 
A sketch of three possible types of particle trajectory in the vicinity of the moonlet is shown in Fig. \ref{fig:trajectory}. 
These correspond to three distinct regimes, a) denoting circulating orbits, b) and c) denoting orbits that result in a 
collision with particles in the vicinity of the moonlet
and directly with the moonlet respectively, and d) denoting horseshoe orbits. All these types of trajectory occur for particles
that are initially in circular orbits, both interior and exterior to the moonlet, when at large distances from it.

Approximating the gravitational force due to the central object by its first order Taylor expansion
about the origin leads to Hill's equation, governing the motion of a particle of mass $m_1$ of the form 
\begin{eqnarray}
\ddot \mathbf{r} &=& -2\Omega \mathbf{e}_z\times \dot \mathbf{r} + 3 \Omega^2 (\mathbf{r} \cdot \mathbf{e}_x) \mathbf{e}_x - \mathbf{\nabla} \Psi/m_1, \label{eq:hills}
\end{eqnarray}
where $\mathbf{r} = (x,y,z)$ is position of a particle with mass $m_1$ and $\Psi$ is the gravitational potential
acting on the particle due to other objects of interest such as the moonlet. 

The square amplitude of the epicyclic motion ${\cal E}^2$ can be defined through
\begin{eqnarray}
{\cal E}^2 = {\Omega^{-2}{\dot x}^2+(2\Omega^{-1}\dot y+3x )^2} \label{eq:defe2}.
\end{eqnarray}
Note that neither the eccentricity $e$, nor the semi-major axis $a$ are formally defined in the local coordinate system. 
However, in the absence of interaction with other masses $(\Psi=0),$ ${\cal E}$ is conserved,
and up to first order in the eccentricity we may make the identification ${\cal E}=e\,a.$ 
We recall the classical definition of the eccentricity $e$ in a coordinate
system centred on the central object
\begin{eqnarray}
e = \left| \frac{\mathbf w \times \left( \mathbf s \times \mathbf w \right)}{\Omega^2a^3} - \frac{\mathbf s}{|\mathbf s |}\right|. \label{eq:defe1}
\end{eqnarray}
Here, $\mathbf{s}$ is the position vector $(a,\,0,\,0)$ and $\mathbf w$ is the velocity vector of the particle relative to the mean shear 
associated with circular Keplerian motion as viewed in the local coordinate system
plus the circular Keplerian velocity corresponding to the orbital frequency $\Omega$ 
of the origin. Thus
\begin{eqnarray}
\mathbf w &=& \left(\dot x,\,\,\dot y + \frac32 \Omega x+ a \Omega,\,\,\dot z\right).
\end{eqnarray}
All eccentricities considered here are 
very small ($\sim 10^{-8}-10^{-7}$) so that the difference between the quantities defined through use of Eqs. \ref{eq:defe2} and \ref{eq:defe1} is negligible, allowing us to adopt $\cal E$ as a measure of the eccentricity throughout this paper.

Let us define another quantity, $\cal A$, that is also conserved for non interacting particle motion, $\Psi=0,$ which
is the $x$ coordinate of the centre of the epicyclic motion and is given by 
\begin{eqnarray}
{\cal A} &=& 2 \Omega^{-1}{\dot y} + 4 x.
\end{eqnarray}
We identify a change in ${\cal A}$ as a change in the semi-major axis $a$ of the particle, again, under the assumption that the eccentricity is small.

\subsection{Two interacting particles}
We now consider the motion of two gravitationally interacting particles with position 
vectors ${\bf r}_1= (x_1,y_1,z_1)$ and ${\bf r}_2=(x_2,y_2,z_2).$ Their corresponding masses are $m_1$ and $m_2$, respectively. 
The governing equations of motion are
\begin{eqnarray}
\ddot \mathbf{r}_1 &=& -2\Omega \mathbf{e}_z\times \dot \mathbf{r}_1 + 3 \Omega^2 (\mathbf{r}_1 \cdot \mathbf{e}_x) \mathbf{e}_x - \mathbf{\nabla}_{\mathbf{r}_1}\Psi_{12}/m_1\\
\ddot \mathbf{r}_2 &=& -2\Omega \mathbf{e}_z\times \dot \mathbf{r}_2 + 3 \Omega^2 (\mathbf{r}_2 \cdot \mathbf{e}_x) \mathbf{e}_x + \mathbf{\nabla}_{\mathbf{r}_1}\Psi_{12}/m_2,
\end{eqnarray}
where the interaction gravitational potential is
$\Psi_{12}= -Gm_1m_2 / \left|\mathbf{r}_1-\mathbf{r}_2\right|$.
The position vector of the centre of mass of the two particles is given by
\begin{eqnarray}
\mathbf{\bar r}&=&\frac{m_1 \mathbf{r}_1 + m_2\mathbf{r}_2 }{m_1+m_2}.
\end{eqnarray}
The vector $\mathbf{\bar r}$ also obeys Eq. \ref{eq:hills} with $\Psi=0$, which applies to an isolated particle. 
This is because the interaction potential does not affect the motion of the centre of mass.
We also find it useful to define the vector
\begin{eqnarray}
{\mbox{\boldmath${\cal E}$}}_i = (\Omega^{-1}{\dot x}_i,\;2\Omega^{-1}\dot y_i+3x_i )\ \ \ \ i=1,2 \label{eq:defevec}.
\end{eqnarray}
Then, consistently with our earlier definition of ${\cal E}$, we have ${\cal E}_i =| {\mbox{\boldmath${\cal E}$}}_i|.$
The amplitude of the epicyclic motion of the centre of mass $\bar {\cal E}$ is given by
\begin{eqnarray}
(m_1+m_2)^2\bar{\cal E}^2 &=& m_1^2{\cal E}_1^2+m_2^2{\cal E}_2^2 + 2m_1m_2{\cal E}_1{\cal E}_2 \cos(\phi_{12}).
\end{eqnarray}
Here $\phi_{12}$ is the angle between $ {\mbox{\boldmath${\cal E}$}}_1$ and $ {\mbox{\boldmath${\cal E}$}}_2.$
It is important to note that $\bar {\cal E}$ is conserved even if the two
particles approach
each other and become bound. This is as long as frictional forces are internal to the
two particle system and do
not affect the centre of mass motion.

\section{Effects leading to damping of the eccentricity of the moonlet}\label{dampe}
We begin by estimating the moonlet eccentricity damping rate associated with ring particles that either collide
directly with the moonlet, particles in its vicinity, or only interact gravitationally with the moonlet.

\subsection{The contribution due to particles impacting the moonlet} \label{sec:edampcollision}
Particles impacting the moonlet in an eccentric orbit exchange momentum with it.
Let us assume that a ring particle, identified with $m_1$ has zero epicyclic amplitude, so that ${\cal E}_1=0$ far away from the moonlet. 
The moonlet is identified with $m_2$ and has an initial epicyclic amplitude ${\cal E}_2$. 
The epicyclic amplitude of the centre of mass is therefore 
\begin{eqnarray}
\bar{\cal E} = \frac{m_2}{m_1+m_2} {\cal E}_2 \simeq \left(1-\frac{m_1}{m_2}\right) {\cal E}_2 \label{eq:ecm},
\end{eqnarray}
where we have assumed that $m_1\ll m_2$. 

The moonlet is assumed to be in a steady state in which there is no net accretion of ring particles. 
Therefore, for every particle that either collides directly with the moonlet or nearby particles bound to it (and so itself becomes temporarily bound to it), one particle must also escape from the moonlet. 
This happens primarily through slow leakage from locations close to the $\mathrm{L}_{1}$ and $\mathrm{L}_{2}$ points 
such that most particles escape from the moonlet with almost zero velocity (as viewed from the centre of mass frame)
and so do not change its orbital eccentricity.
However, after an impacting particle becomes bound to the moonlet, conservation of the epicyclic amplitude associated
with the centre of mass motion together with Eq. \ref{eq:ecm} imply that
each impacting particle will reduce the eccentricity by a factor $1-m_1/m_2.$ 

It is now an easy task to estimate the eccentricity damping timescale by determining the number of particle
collisions per time unit with the moonlet or particles bound to it. 
To do that, a smooth window function $W_{b+c}(b)$ is used, being unity 
for impact parameters $b$ that always result in an impact with the moonlet or particles nearby that are bound to it, being zero for impact parameters that never result in an impact. 

The number of particles impacting the moonlet per time unit, $dN/dt$, is obtained by integrating over the impact parameter
with the result that
\begin{eqnarray}
\frac{dN}{dt} &=& \frac{1}{m_1}{\int_{-\infty}^{\infty} \frac32 \, \Sigma\, \Omega \,|b| \; W_{b+c}(b)\, \mathrm{d}b}\label{Window}.
\end{eqnarray}
We note that allowing $b$ to be negative enables impacts from both sides of the moonlet to be taken into account.

Therefore, after using Eq. \ref{eq:ecm} we find that the rate of change of the moonlet's eccentricity $e_2$, or equivalently of it's amplitude of epicyclic motion ${\cal E}_2$, is given by
\begin{eqnarray}
\frac{d{\cal E}_2}{dt} = -\frac{{\cal E}_2}{\tau_{e,\mathrm{collisions}}} &=& -\frac{{\cal E}_2}{m_2}{\int_{-\infty}^{\infty} \frac32 \, \Sigma\, \Omega \,|b| \; W_{b+c}(b)\, \mathrm{d}b}, \label{eq:d13}
\end{eqnarray}
where $\tau_{e,\mathrm{collisions}}$ defines the circularisation time arising from collisions with the moonlet. We remark that the natural unit
for $b$ is the Hill radius of the moonlet, $r_H = (m_2/(3M_p))^{1/3}a,$ so that the dimensional scaling for
$\tau_{e,\mathrm{collisions}}$ is given by 
\begin{eqnarray}
\tau_{e,\mathrm{collisions}}^{-1} &\propto& G\;\Sigma\; r_H^{-1}\;\Omega^{-1} \label{eq:scaling}
\end{eqnarray}
which we find to also apply to all the processes for modifying the moonlet's eccentricity discussed below.
If we assume that $W_{b+c}(|b|)$ can be approximated by a box function, being unity in the interval $[1.5r_H,\,2.5r_H]$ and zero elsewhere, we get
\begin{eqnarray}
\tau_{e,\mathrm{collisions}}^{-1} &=& 2.0 \;G\;\Sigma\; r_H^{-1}\;\Omega^{-1} \label{eq:taucol}.
\end{eqnarray}

\subsection{Eccentricity damping due the interaction of the moonlet with particles on horseshoe orbits}\label{sec:edamphorseshoe}
The eccentricity of the moonlet manifests itself in a small oscillation of the moonlet about the origin.
Primarily ring particles on horseshoe orbits will respond to that oscillation and damp it.
This is because only those particles on horseshoe orbits spend enough time in the vicinity of the moonlet, i.e. many epicyclic periods.

In appendix \ref{app:response}, we calculate the amplitude of epicyclic motion ${\cal E}_{1f}$ (or equivalently the eccentricity $e_{1f}$) that is induced in a single ring particle in a horseshoe orbit undergoing a close approach to a moonlet which is assumed to be in an eccentric orbit. 
In order to calculate the circularisation time, we have to consider all relevant impact parameters. 
We begin by noting that each particle encounter with the moonlet 
is conservative and is such that for each particle,
the Jacobi constant, applicable when the moonlet is in circular orbit, is increased
by an amount $m_1\Omega^2{\cal E}_{1f}^2/2$ by the action of
the perturbing force, associated with the eccentricity of the moonlet, as the particle passes by.
Because the Jacobi constant, or energy in the rotating frame, for the moonlet and the particle together is conserved,
the square of the epicyclic amplitude associated with the moonlet alone changes by ${\cal E}_{1f}^2\,m_1/m_2.$
Accordingly the change in the amplitude of epicyclic motion of the moonlet ${\cal E}_2$, consequent
on inducing the amplitude of epicyclic motion ${\cal E}_{1f}$ in the horseshoe particle, is given by 
\begin{eqnarray}
\Delta {\cal E}_{2}^2 &=& -\frac{m_1}{m_2} {\cal E}_{1f}^2.
\end{eqnarray}
Note that this is different compared to Eq. \ref{eq:ecm}. 
Here, we are dealing with a second order effect. 
First, the eccentric moonlet excites eccentricity in a ring particle.
Second, because the total epicyclic motion is conserved, the epicyclic motion of the moonlet is reduced. 

Integrating over the impact parameters associated with ring particles
and taking into account particles streaming by the moonlet from both directions
by allowing negative impact parameters gives
\begin{eqnarray}
\left.\frac{d{\cal E}_2^2}{dt}\right|_{\mathrm{horseshoe}}&=& -\int_{-\infty}^{\infty} \frac{3\Sigma {\cal E}_{1f}^2 \Omega \,|b| \,W_d(b)\,db}{2m_2} \equiv -\frac{2{\cal E}_2^2}{\tau_{e,\mathrm{horseshoe}}},
\end{eqnarray}
where $\tau_{e,\mathrm{horseshoe}}$ is the circularisation time and, as above, we have inserted a window function, which
is unity on impact parameters that lead to horseshoe orbits, otherwise being zero.
Using ${\cal E}_{1f}$ given by Eq. \ref{inducede}, we obtain
\begin{eqnarray}
\tau_{e,\mathrm{horseshoe}}^{-1} =\frac{9}{128}\left(\frac{\Sigma r_H^2\Omega}{m_2}\right)\int_{-\infty}^{\infty}{\cal I}^2\eta^{4/3} W_d\left((2\eta r_h)^{1/6}\right)\,d\eta.
\end{eqnarray}
For a sharp cutoff of $W_d(b)$ at $b_m =1.5r_H$ we find the value of the integral in the above to be $2.84$.
Thus, in this case we get 
\begin{eqnarray}
\tau_{e,\mathrm{horseshoe}}^{-1}= 0.13 \;G \;\Sigma\;r_H^{-1}\;\Omega^{-1}.\label{eq:damphorse}
\end{eqnarray}
However, note that this value is sensitive to the value of $b_m$ adopted. For $b_m= 1.25r_H,$ $t_c$ is a factor of $4.25$ larger.

\subsection{The effects of circulating particles}\label{sec:dampecirc}
The effect of the response of circulating particles to the gravitational
perturbation of the moonlet on the moonlet's eccentricity can be estimated from the work of 
\cite{GoldreichTremaine1980} and \cite{GoldreichTremaine1982}.
These authors considered a ring separated from a satellite such that co-orbital effects
were not considered. Thus, their expressions may be applied to estimate
effects due to circulating particles. However, we exclude their corotation torques as they 
are determined by the ring edges and are absent in a local model with uniform azimuthally averaged surface density.
Equivalently, one may simply assume that the corotation torques are saturated.
When this is done only Lindblad torques act on the moonlet. These tend to excite the moonlet's eccentricity rather than damp it.

We replace the ring mass in Eq. 70 of \cite{GoldreichTremaine1982} by the integral over the impact parameter and switch to our notation to obtain
\begin{eqnarray}
\left.\frac{d{\cal E}_2^2}{dt}\right|_{\mathrm{circ}}&=& 9.55 \int_{-\infty}^{\infty} m_2 \, \Sigma \, G^2 \, \Omega^{-3} \,  |b|^{-5} \;{\cal E}_2^2\;W_a(b) \,\mathrm{d}b,\label{Goldcirc}
\end{eqnarray}
where $W_a(b)$ is the appropriate window function for circulating particles. 
Assuming a sharp cutoff of $W_a(|b|)$ at $b_m$, we can evaluate the integral in Eq. \ref{Goldcirc} to get 
\begin{eqnarray}
\frac{1}{{\cal E}_2}\left.\frac{d{\cal E}_2}{dt}\right|_{\mathrm{circ}}=-\frac{1}{\tau_{e,circ}}&=&  0.183 \;G \;\Sigma\;r_H^{-1}\;\Omega^{-1},
\label{Goldcirc2} \end{eqnarray}
where we have adopted $b_m=2.5r_H$, consistently with the simulation results presented below.
We see that, although $\tau_{e,circ}<0$ corresponds to growth rather than damping of the eccentricity, it scales in the same manner
as the circularisation times in Sects. \ref{sec:edampcollision} and \ref{sec:edamphorseshoe} scale (see Eq. \ref{eq:scaling}). 

This timescale and the timescale associated with particles on horseshoe orbits, $\tau_{e,\mathrm{horseshoe}}$, are significantly larger than the timescale associated with collisions, given in Eq. \ref{eq:taucol}.
We can therefore ignore those effects for most of the discussion in this paper.

\section{Processes leading to the excitation of the eccentricity of the moonlet} \label{sec:excitee}
In Sec. \ref{dampe} we assumed that the moonlet had a small eccentricity but neglected
the initial eccentricity of the impacting ring particles.
When this is included, collisions and gravitational interactions of ring particles with the moonlet may also excite its eccentricity.

\subsection{Collisional eccentricity excitation}
To see this, assume that the moonlet initially has zero eccentricity,
or equivalently no epicyclic motion, but the ring particles do.

As above we consider the conservation of the epicyclic motion of the centre of mass in
order to connect the amplitude of the
final epicyclic motion of the combined moonlet and ring particle
to the initial amplitude of the ring particle's epicyclic motion.
This gives the epicyclic amplitude of the centre of mass after one impact as
\begin{eqnarray}
\bar{\cal E} = \frac{m_1}{m_1+m_2} {\cal E}_1 \simeq \left(\frac{m_1}{m_2}\right) {\cal E}_1 \label{eq:ecmexc}.
\end{eqnarray}
Assuming that successive collisions are uncorrelated and occur stochastically
with the mean time between consecutive encounters being $\tau_{ce}$, the evolution of $\bar{\cal E}$ is governed by the equation
\begin{eqnarray}
\left.\frac{d \bar{\cal E}^2}{dt}\right|_{\mathrm{collisions}} = \left( \frac{m_1}{m_2}\right)^{2} \langle {\cal E}_1^{2} \rangle \tau_{ce}^{-1},\label{eq:edot}
\end{eqnarray}
where $\tau_{ce}^{-1} = dN/dt$ can be expressed in terms of the surface density and an impact window function $W_{b+c}(b)$ (see Eq. \ref{Window}).
The quantity $\langle {\cal E}_1^{2} \rangle$ is the mean square value of ${\cal E}_1$ for the ring particles.
Using Eq. \ref{eq:defe2}, this may be written in terms of the mean squares of the components of the velocity dispersion
relative to the background shear, in the form
\begin{eqnarray}
\langle {\cal E}_1^{2} \rangle = \Omega^{-2}\langle ({\dot x_1}^2+4({\dot y_1}+3\Omega x_1/2 )^2\rangle .
\end{eqnarray}
We find in numerical simulations $\langle e_1\rangle\sim10^{-6}$ for almost all ring parameters. This value is mainly determined by the coefficient of restitution \citep[see e.g. Fig 4 in][]{MorishimaSalo2006}.

\subsection{Stochastic excitation due to circulating particles}\label{circpart}
Ring particles that are on circulating orbits, such as that given by path (a) in Fig. \ref{fig:trajectory}, exchange energy and angular momentum with the moonlet and therefore change its eccentricity. 
\cite{GoldreichTremaine1982} calculated the change in the eccentricity of a ring particle due to a satellite. 
We are interested in the change of the eccentricity of the moonlet induced by a ring of particles and therefore swap the satellite mass with the mass of a ring particle.
Thus, rewriting their results (Eq. 64 in \cite{GoldreichTremaine1982}) in our local notation without reference to the semi-major axis, we have
\begin{eqnarray}
\Delta {\cal E}_2^2 &=& 5.02 \;m_1^2\;G^2\;\Omega^{-4}\; {b}^{-4}.
\end{eqnarray}
Supposing that the moonlet has very small eccentricity, it will receive stochastic impulses that cause its eccentricity to undergo
a random walk that will result in ${\cal E}_2^2$ increasing linearly with time, so that we may write
\begin{eqnarray}
\left.\frac{d{\cal E}_2^2}{dt}\right|_{\mathrm{circulating\;particles}} 
&=&
\int_{-\infty}^{\infty} W_a(b) \Delta {\cal E}_2^2 d(1/t_b),
\end{eqnarray}
where $d(1/t_b)$ is the mean encounter rate with particles which have an impact parameter in the interval $(b,b+db)$.
$W_a(b)$ is the window function describing the band of particles in circulating orbits.
Setting $d(1/t_b)=3\Sigma\Omega |b| db/(2m_1),$ we obtain
\begin{eqnarray}
\left.\frac{d{\cal E}_2^2}{dt}\right|_{\mathrm{circulating\;particles}} 
&=& 7.53\int_{-\infty}^{\infty} W_a(b) m_1 |b|^{-3}\Sigma G^2 \Omega^{-3}\mathrm{d}b.
\label{circpe}
\end{eqnarray}
For high surface densities an additional effect can excite the eccentricity when gravitational wakes occur.
The Toomre parameter $Q$ is defined as 
\begin{eqnarray}
Q=\frac{\bar{v} \Omega}{\pi G\Sigma},
\end{eqnarray}
where $\bar v$ is the velocity dispersion of the ring particles\footnote{The Toomre criterion used here was originally derived for a flat gaseous disc.
To make use of it we replace the sound speed by the radial velocity dispersion of the ring particles.}. 
When the surface density is sufficiently high such that $Q$ approaches unity, transient clumps will form in the rings. 
The typical length scale of these structures is given by the critical Toomre wave length $\lambda_\mathrm{T} = {2 \pi^2 G \Sigma}{\Omega^{-2}}$ \citep{Daisaka2001}, which can be used to estimate a typical mass of the clump:
\begin{eqnarray}
m_\mathrm{T} 	&\sim& \pi \lambda_\mathrm{T}^2 \Sigma
	= 	{4\pi^5G^2\Sigma^3}{\Omega^{-4}}.
\end{eqnarray}
Whenever strong clumping occurs, $m_\mathrm{T}$ should be used in the above calculation instead of the mass of a single particle $m_1$:
\begin{eqnarray}
\left.\frac{d{\cal E}_2^2}{dt}\right|_{\mathrm{circulating\;clumps}}&=& 7.53 \int_{-\infty}^{\infty} W'_a(b) m_T |b|^{-3}\Sigma G^2 \Omega^{-3} \mathrm{d}b,
\label{clumpe}
\end{eqnarray}
where $W'_a(b)\approx W_a(b)$ is the appropriate window function.
For typical parameters used in Sect. \ref{sec:results}, this transition occurs at $\Sigma\sim200\mathrm{kg/m^2}$.

\subsection{Equilibrium eccentricity}
Putting together the results from this and the previous section, we can estimate an equilibrium eccentricity of the moonlet. 
Let us assume the eccentricity $e_2$, or the amplitude of the epicyclic motion ${\cal E}_2$, evolves under the influence of excitation and damping forces as follows
\begin{eqnarray}
\frac{d{\cal E}_2^2}{dt}&=& -2 \left(
 \tau_{e,\mathrm{collisions}}^{-1} 
 + \tau_{e,\mathrm{horseshoe}}^{-1} 
 +\tau_{e,\mathrm{circ}}^{-1} \right) {\cal E}_2^2 \nonumber \\
&&+\left.\frac{d{\cal E}_2^2}{dt}\right|_{\mathrm{collisions}}\nonumber
 +\left.\frac{d{\cal E}_2^2}{dt}\right|_{\mathrm{circulating\;particles}}\\
&&+\left.\frac{d{\cal E}_2^2}{dt}\right|_{\mathrm{circulating\;clumps}}
.\label{equile}
\end{eqnarray}
The equilibrium eccentricity is then found by setting the above equation equal to zero and solving for ${\cal E}_2$. 

To make quantitative estimates we need to specify
the window functions $W_a(b)$, $W_{b+c}(b)$ and $W_d(b)$ that determine in which impact parameter
bands the particles are in (see Fig. \ref{fig:trajectory}). 
To compare our analytic estimates to the numerical simulations presented below, we measure the window functions numerically.
Alternatively, one could simply use sharp cutoffs at some multiple of the Hill radius. We already made use of this approximation as a simple estimate in the previous sections. 
The results may vary slightly, but not significantly. 

However, as the window functions are dimensionless, it is possible to obtain the dimensional scaling of ${\cal E}_2$ by adopting the length scale applicable to the impact parameter to be the Hill radius $r_H$ and simply assume that the window functions are of order unity. 
As already indicated above, all of the circularisation times scale as $\tau_{e} \propto \Omega \,r_H\,G^{-1}\,\Sigma^{-1}$, or equivalently $\propto m_2/(\Sigma r_H^2\Omega)$.
Assuming the ring particles have zero velocity dispersion, the eccentricity excitation is then due to circulating clumps or particles and the scaling of $d{\cal E}_2^2/{dt}$ due to this cause is given by Eqs. \ref{circpe} and \ref{clumpe} by
\begin{eqnarray}
\frac{d{\cal E}_2^2}{dt}&\propto& m_i r_H^{-2}\Sigma G^2 \Omega^{-3},
\end{eqnarray}
where $m_i$ is either $m_1$ or $m_T.$ We may then find the scaling
of the equilibrium value of ${\cal E}_2$ from consideration of Eqs. \ref{equile} and \ref{eq:d13} as
\begin{eqnarray}
{\cal E}_2&\propto & \frac{m_i}{m_2} r_H^{2}.
\end{eqnarray}
This means that the expected kinetic energy in the non circular motion of the moonlet
is $\propto m_i r_H^{2}\Omega^2$ which can be viewed as stating that
the non circular moonlet motion scales in equipartition with the mass $m_i$ moving 
with speed $r_H\Omega.$ This speed applies when the dispersion velocity associated 
with these masses is zero indicating that the shear across a Hill radius
replaces the dispersion velocity in that limit. 

In the opposite limit when the dispersion velocity exceeds the shear across a Hill
radius and the dominant source of eccentricity excitation is due to collisions,
Eq. \ref{equile} gives 
\begin{eqnarray}
m_2{\cal E}_2&=& m_1\langle {\cal E}_1^{2} \rangle 
\end{eqnarray}
so that the moonlet is in equipartition with the ring particles.
Results for the two limiting cases can be combined to give
an expression for the amplitude of the epicyclic motion excited in the moonlet
of the form
\begin{eqnarray}
m_2\Omega^2{\cal E}_2&=& m_1\Omega^2 \langle {\cal E}_1^{2} \rangle + C_im_i\Omega^2 r_H^{2},
\end{eqnarray}
where $C_i$ is a constant of order unity.
This indicates the transition between the shear dominated and the velocity dispersion
dominated limits.

\section{Processes leading to a random walk in the semi-major axis of the moonlet}\label{sec:randa}
We have established estimates for the equilibrium eccentricity of the moonlet in the previous section. 
Here, we estimate the random walk of the semi-major axis of the moonlet. 
In contrast to the case of the eccentricity, there is no tendency
for the semi-major axis to relax to any particular value, so that there are no damping terms
and the deviation of the semi-major axis from its value at time $t=0$
grows on average as $\sqrt{t}$ for large $t.$

Depending on the surface density, there are different effects that dominate the contributions
to the random walk of the moonlet. For low surface densities, collisions and horseshoe orbits are most important.
For high surface densities, the random walk is dominated by the stochastic gravitational force of
circulating particles and clumps. In this section, we try to estimate the strength of each effect.

\subsection{Random walk due to collisions}\label{sec:randacol}
Let us assume, without loss of generality, that the guiding centre of the epicyclic motion of a ring particle is displaced from the
orbit of the moonlet by ${\cal A}_1$ in the inertial frame, whereas in the co-rotating frame the moonlet is initially located at the origin with ${\cal A}_2=0$. 
When the impacting particle becomes bound to the moonlet, the guiding centre of the epicyclic motion of the centre of mass of the combined object, as viewed in the inertial
frame, is then displaced from the original
moonlet orbit by 
\begin{eqnarray}
\Delta \bar{\cal A} = \frac{m_1}{m_1+m_2} {\cal A}_1 \sim \frac{m_1}{m_2} {\cal A}_1 .
\end{eqnarray}
This is the analogue of Eq. \ref{eq:ecmexc} for the evolution of the semi-major axis. For an
ensemble of particles with impact parameter $b,$ 
the average centre of epicyclic motion is $\langle {\cal A}_1\rangle=b$. Thus we can write the evolution of $\bar{\cal A}$ due to consecutive encounters as
\begin{eqnarray}
\left.\frac{d\bar{\cal A}^2}{dt}\right|_{\mathrm{collisions}} &=& \left(\frac{m_1}{m_2}\right)^2 \langle {\cal A}_1^2\rangle\; \tau_{ce}^{-1}\\
&=& \frac{m_1}{m_2^2}\;\int_{-\infty}^\infty \frac32\, \Sigma\, \Omega \;\left|b\right|^3\; W_{b+c}(b) \, \mathrm{d}b, \label{eq:randwalka:collisions}
\end{eqnarray}
which should be compared to the corresponding expression for the eccentricity in Eq. \ref{eq:edot}.

\subsection{Random walk due to stochastic forces from circulating particles and clumps}
Particles on a circulating trajectory (see path (a) in Fig. \ref{fig:trajectory}) that come close the moonlet will exert a stochastic gravitational force.
The largest contributions occur for particles within a few Hill radii.
Thus, we can crudely estimate the magnitude of the specific gravitational force (acceleration) due to a single particle as
\begin{eqnarray}
f_{cp} &=& \frac{G\,m_1}{(2r_H)^2}.
\end{eqnarray}
When self-gravity is important, gravitational wakes (or clumps) have to be taken into account, as was done in Sec. \ref{circpart}. 
Then a rough estimate of the largest specific gravitational forces due to self gravitating clumps is 
\begin{eqnarray}
f_{cc} &=& \frac{G\,m_\mathrm{T}}{\left(2r_H+\lambda_T\right)^2},
\end{eqnarray}
where $2r_H$ has been replaced by $2r_H+\lambda_T$. This allows for the fact 
that $\lambda_T$ could be significantly larger than $r_H$, in which case the approximate distance of the clump to the moonlet is $\lambda_T$. 

Following the formalism of \cite{ReinPapaloizou2009}, we define a diffusion coefficient as 
$D= 2\tau_c \langle f^2\rangle,$ where $f$ is an acceleration, $\tau_c$ is the correlation time
and the angle brackets denote a mean value.
We here take the correlation time associated with
these forces to be the orbital period, $2\pi\Omega^{-1}.$
This is the natural dynamical timescale of the systems and has been found to be a reasonable assumption from an analysis of the simulations described below.
By considering the rate of change of the energy of the moonlet motion, we may estimate the random walk in $\bar {\cal A}$ due to circulating particles and self-gravitating clumps 
to be given by \citep{ReinPapaloizou2009}:
\begin{eqnarray}
\left.\frac{d\bar{\cal A}^2}{dt}\right|_{\mathrm{circulating\;particles}} &=& 4 D_{cp}\,\Omega^{-2}
= 16\pi\, \Omega^{-3} \langle f_{cp}^2\rangle \\
&=&16 \pi \, \Omega^{-3}\left \langle \left(\frac{G\,m_1}{(2r_H)^2}\right)^2\right \rangle\label{eq:randwalka:circulatingparticles}\\
\left.\frac{d\bar{\cal A}^2}{dt}\right|_{\mathrm{circulating\;clumps}} &=& 4 D_{cc}\,\Omega^{-2}
= 16\pi\, \Omega^{-3} \langle f_{cc}^2\rangle\\
&=&16 \pi\,\Omega^{-3}\left\langle\left(
\frac{G\,m_\mathrm{T}}{\left(2r_H+\lambda_T\right)^2}\right)^2\right\rangle. \label{eq:randwalka:circulatingclumps}
\end{eqnarray}
This is only a crude estimate of the random walk undergone by the moonlet. In reality several additional effects might also play a role. For example, circulating particles and clumps are clearly correlated, the gravitational wakes have a large extent in the azimuthal direction and particles that spend a long time in the vicinity of the moonlet have more complex trajectories. 
Nevertheless, we find that the above estimates are correct up to a factor 2 for all the simulations that we performed (see below).

\subsection{Random walk due to particles in horseshoe orbits}
Finally, let us calculate the random walk induced by particles on horseshoe orbits.
Particles undergoing horseshoe turns on opposite sides of the planet produce changes of opposite sign.
Encounters with the moonlet are stochastic and therefore the semi-major axis will undergo a random walk.
A single particle with impact parameter $b$ will change the semi-major axis of the moonlet by
\begin{eqnarray}
\Delta\bar{\cal A} = 2 \frac{m_1}{m_1+m_2} b.
\end{eqnarray}
Analogous to the analysis in Sec. \ref{sec:randacol}, the time evolution of $\bar {\cal A}$ is then governed by
\begin{eqnarray}
\left.\frac{d\bar{\cal A}^2}{dt}\right|_{\mathrm{horseshoe}} &=& 4 \left(\frac{m_1}{m_2}\right)^2 b^2\; \tau_{he}^{-1}\\
&=& 6 \frac{m_1}{m_2^2}\;\int_{-\infty}^\infty \Sigma\, \Omega \;\left|b\right|^3\; W_d(b) \, \mathrm{d}b.\label{eq:randwalka:horseshoe}
\end{eqnarray}
Note that this equation is identical to Eq. \ref{eq:randwalka:collisions} except a factor 4, as particles with impact parameter $b$ will leave the vicinity of the moonlet at $-b$.

\section{Numerical Simulations}\label{sec:numerical}
\begin{table*}[tb]
\begin{center}
\begin{tabular}{l|rr|rrr}
Name 		&	\multicolumn{1}{c}{$\Sigma$} 		& \multicolumn{1}{c|}{$r_2$} 		& \multicolumn{1}{c}{$dt$} & \multicolumn{1}{c}{$L_x\times L_y$} 				& \multicolumn{1}{c}{$N$} \\\hline\hline
\texttt{EQ5025}	& 	$50\,\mathrm{kg/m^2}$	& $25\,$m			& 4s& $1000\,\mathrm{m}\times1000\,\mathrm{m}$	& 7.2k \\
\texttt{EQ5050}	& 	$50\,\mathrm{kg/m^2}$	& $50\,$m			& 4s& $1000\,\mathrm{m}\times1000\,\mathrm{m}$	& 7.2k \\
\texttt{EQ20025}& 	$200\,\mathrm{kg/m^2}$	& $25\,$m			& 4s& $1000\,\mathrm{m}\times1000\,\mathrm{m}$	& 28.8k \\
\texttt{EQ20050}& 	$200\,\mathrm{kg/m^2}$	& $50\,$m			& 4s& $1000\,\mathrm{m}\times1000\,\mathrm{m}$	& 28.8k \\
\texttt{EQ40025}& 	$400\,\mathrm{kg/m^2}$	& $25\,$m			& 4s& $1000\,\mathrm{m}\times1000\,\mathrm{m}$	& 57.6k \\
\texttt{EQ40050}& 	$400\,\mathrm{kg/m^2}$	& $50\,$m			& 4s& $1000\,\mathrm{m}\times1000\,\mathrm{m}$	& 57.6k \\
\texttt{EQ40050DT}& 	$400\,\mathrm{kg/m^2}$	& $50\,$m			& 40s& $1000\,\mathrm{m}\times1000\,\mathrm{m}$	& 57.6k \\
\texttt{EQ5050DTW}& 	$50\,\mathrm{kg/m^2}$	& $50\,$m			& 40s& $2000\,\mathrm{m}\times2000\,\mathrm{m}$	& 28.8k \\
\texttt{EQ20050DTW}& 	$200\,\mathrm{kg/m^2}$	& $50\,$m			& 40s& $2000\,\mathrm{m}\times2000\,\mathrm{m}$	& 115.2k \\
\texttt{EQ40050DTW}& 	$400\,\mathrm{kg/m^2}$	& $50\,$m			& 40s& $2000\,\mathrm{m}\times2000\,\mathrm{m}$	& 230.0k \\
\end{tabular}
\caption{Initial simulation parameters. The second column gives the surface density of the ring. The third gives the moonlet radius. The fourth column gives the time step. The fifth and sixth columns give the lengths of the main box as measured in the $xy$-plane
and the number of particles used, respectively. \label{tab:sim}}
\end{center}
\end{table*}

\begin{figure}
\center
\scalebox{0.75}{
\input{boxes.pstex_t}
}
	\caption{Shearing box, simulating a small patch 
	 of a planetary ring system. 
	The particles that leave an auxiliary box in the azimuthal ($y$) direction reenter the same box on the other side and get copied into the main box at the corresponding location. 
	 All auxiliary boxes are equivalent and
	there are no curvature terms present in the shearing sheet.
	\label{fig:boxes}}
\end{figure}
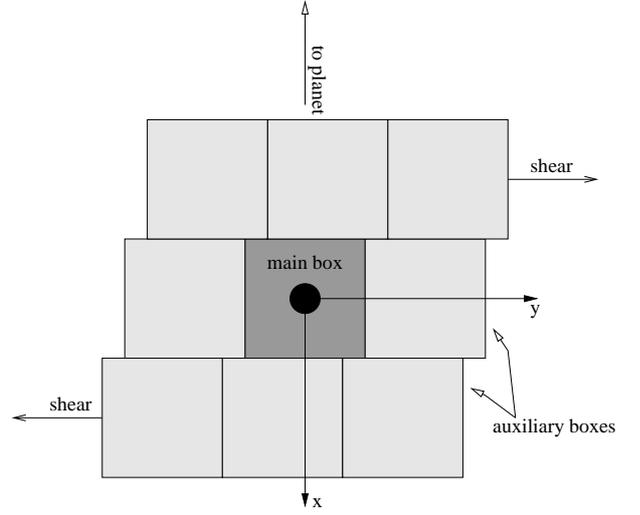

We perform realistic three dimensional simulations of Saturn's rings with an embedded moonlet and verify the analytic estimates presented above. 
The nomenclature and parameters used for the simulations are listed in Table \ref{tab:sim}.

\subsection{Methods} \label{sec:methods}
The gravitational forces are calculated with a Barnes-Hut tree \citep{Barnes1986}. 
The numerical scheme is similar to that used by \cite{ReinLesurLeinhardt2010}. 
Additionally, we implemented a symplectic integrator for Hill's equations \citep{Quinn2010}. 
This turned out to be beneficial for accurate energy conservation at almost no additional cost 
when the eccentricity of the moonlet ($\sim10^{-8}$) was small and integrations were undertaken over many hundreds of dynamical timescales. 

To further speed up the calculations, we run two coupled simulations in parallel. 
The first adopts the main box which incorporates a moonlet. The second adopts an auxiliary box
which initially has the same number of particles but which does not contain a moonlet.
This is taken to be representative of the unperturbed background ring.
We describe this setup as enabling us to adopt \textit{pseudo} shear periodic boundary conditions. 
We consider the main box together with eight equivalent auxiliary boxes to be stacked as illustrated in Fig. \ref{fig:boxes}.
All auxiliary boxes are identical copies whose centres are shifted according to the background shear as in a normal shearing sheet.
On account of this motion auxiliary boxes are removed when their centres are shifted in azimuth by more than $1.5L_y$ from the centre
of the main box and then reinserted in the same orbit on the opposite side of the main box so that the domain under consideration is
prevented from shearing out. 
If a particle in the main box crosses one of its boundaries, it is discarded.
If a particle in the auxiliary box crosses one of its boundaries, it is reinserted on the other side of this box,
according to normal shear periodic boundary conditions.
But in addition it is also copied into the corresponding location in the main box.
We describe this procedure as applying \textit{pseudo} shear periodic boundary conditions.

In a similar calculation, \cite{LewisStewart2009} use a very long box (about 10 times longer than the boxes used in this paper) 
to ensure that particles are completely randomised between encounters with the moonlet. 
We are not interested in the long wavelength response that is created by the moonlet. 
Effects that are most important for the moonlet's dynamical evolution are found to happen within a few Hill radii. 
Using the pseudo shear periodic boundary conditions, we ensure that incoming particles are uncorrelated and do not contain
prior information about the perturber.
This setup speeds up our calculations by more than an order of magnitude.

The gravity acting on a particle in the main box, which also contains the moonlet, 
is calculated by summing over the particles in the main box and all auxiliary boxes. 
The gravity acting on a particle in the auxiliary box is calculated the standard way, 
by using ghost boxes which are identical copies of the auxiliary box.
Tests have indicated that our procedure does not introduce
unwanted fluctuations in the gravitational forces.

The moonlet is allowed to move freely in the main box. 
However, in order to prevent it leaving the computational domain, as
soon as the moonlet has left the innermost part (defined as extending one eighth of the box size), 
all particles are shifted together with the box boundaries, such that the moonlet is returned to the centre of the box. 
This is possible because the shearing box approximation is invariant with respect to translations in the $xy$ plane (see Eq. \ref{eq:hills}). 

Collisions between particles are resolved using the instantaneous collision model and a 
velocity dependent coefficient of restitution given by \cite{Bridges1984}:
\begin{eqnarray}
\epsilon(v) &= &\mathrm{min}\left[0.34\cdot \left( \frac{v_\parallel}{1\mathrm{cm/s}} \right)^{-0.234}, 1\right],
\end{eqnarray}
where $v_\parallel$ is the impact speed projected on the axis between the two particles. 
The already existing tree structure is reused to search for nearest neighbours \citep{ReinLesurLeinhardt2010}.

\subsection{Initial conditions and tests}

\begin{figure}[tb]
\centering
\includegraphics[angle=90,width=0.99\columnwidth]{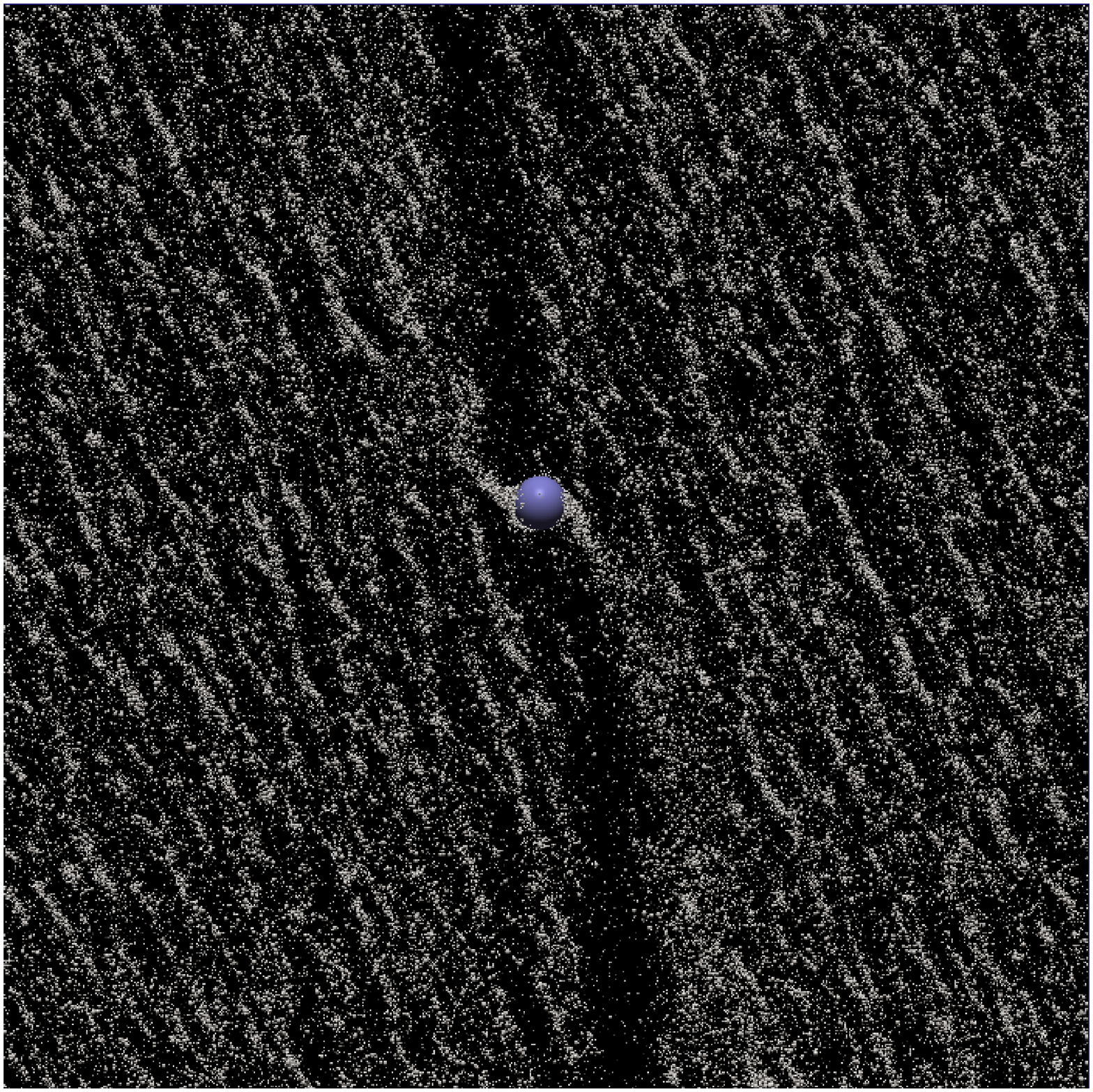}
\caption{Snapshots of the particle distribution and the moonlet (blue) in simulation \texttt{EQ40050DTW}. 
The wake is much clearer than in Fig. \ref{fig:snap} as the box size is twice as large. \label{fig:snapdtw}}
\end{figure}

\begin{figure*}[p]
\centering
\vspace{2cm}
\subfigure[$\Sigma=200\,\mathrm{kg/m^2}$, $r_2=25\,\mathrm{m}$]{
\includegraphics[angle=90,width=0.999\columnwidth]{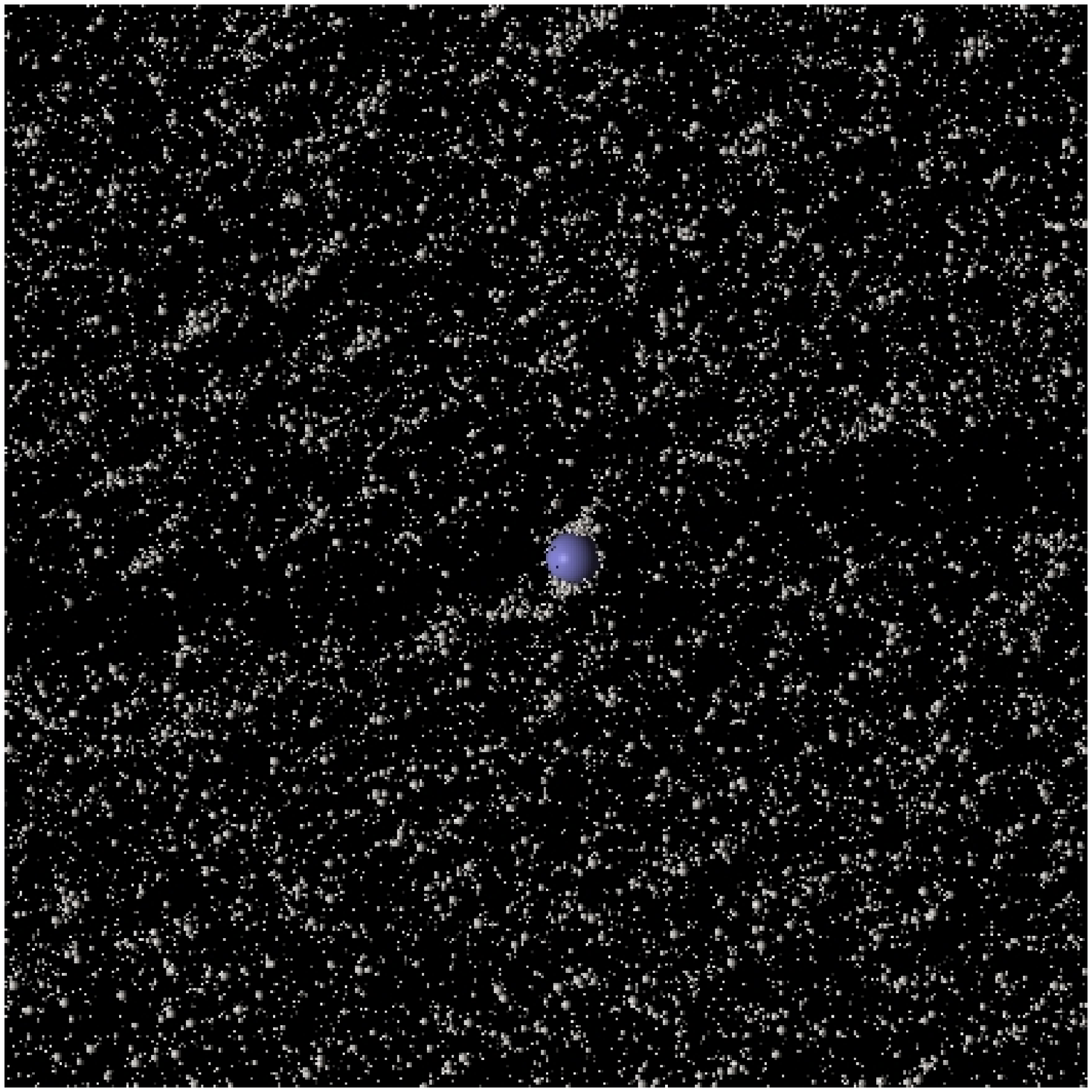}
}
\subfigure[$\Sigma=200\,\mathrm{kg/m^2}$, $r_2=50\,\mathrm{m}$]{
\includegraphics[angle=90,width=0.999\columnwidth]{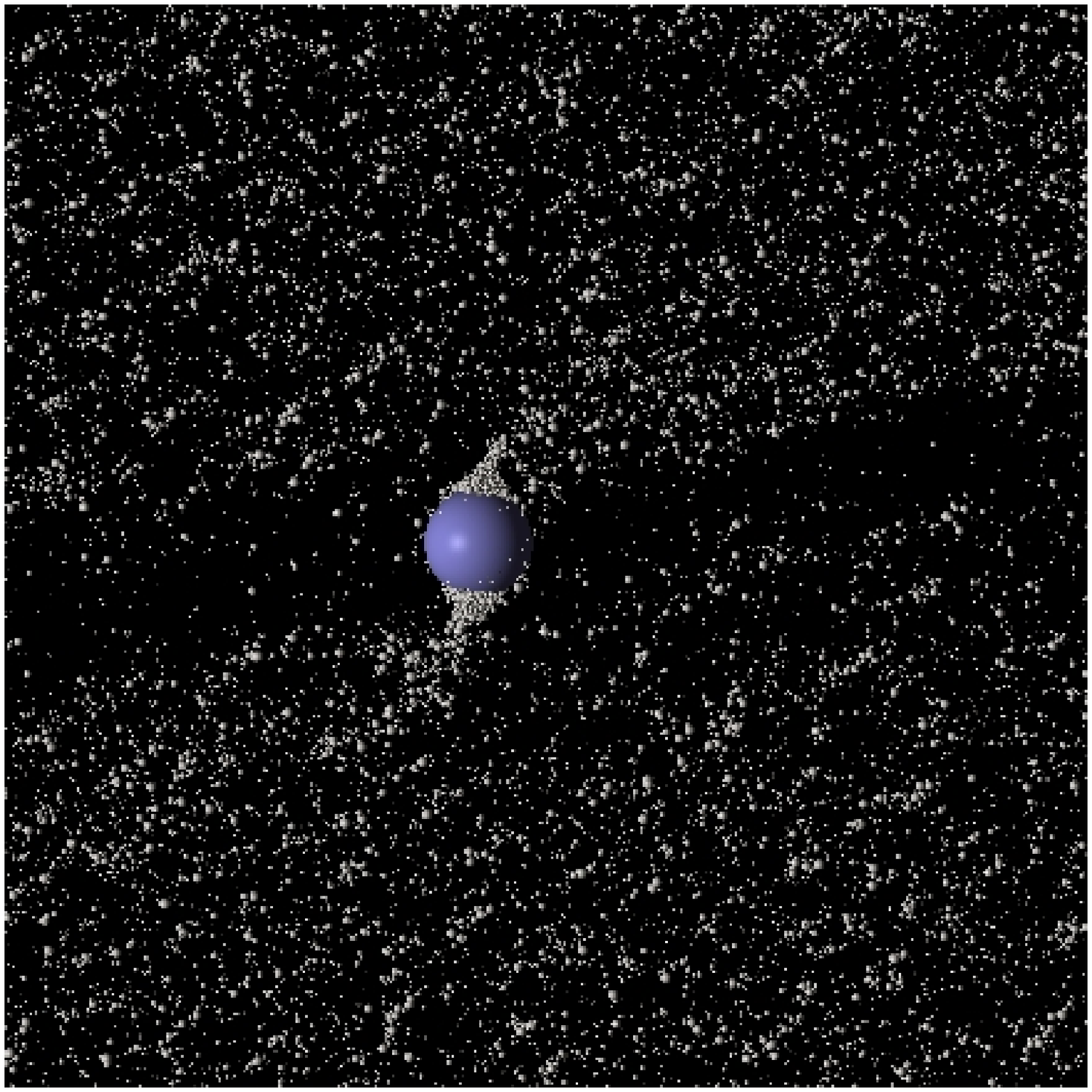}
}
\subfigure[$\Sigma=400\,\mathrm{kg/m^2}$, $r_2=25\,\mathrm{m}$]{\label{fig:snap1c}
\includegraphics[angle=90,width=0.999\columnwidth]{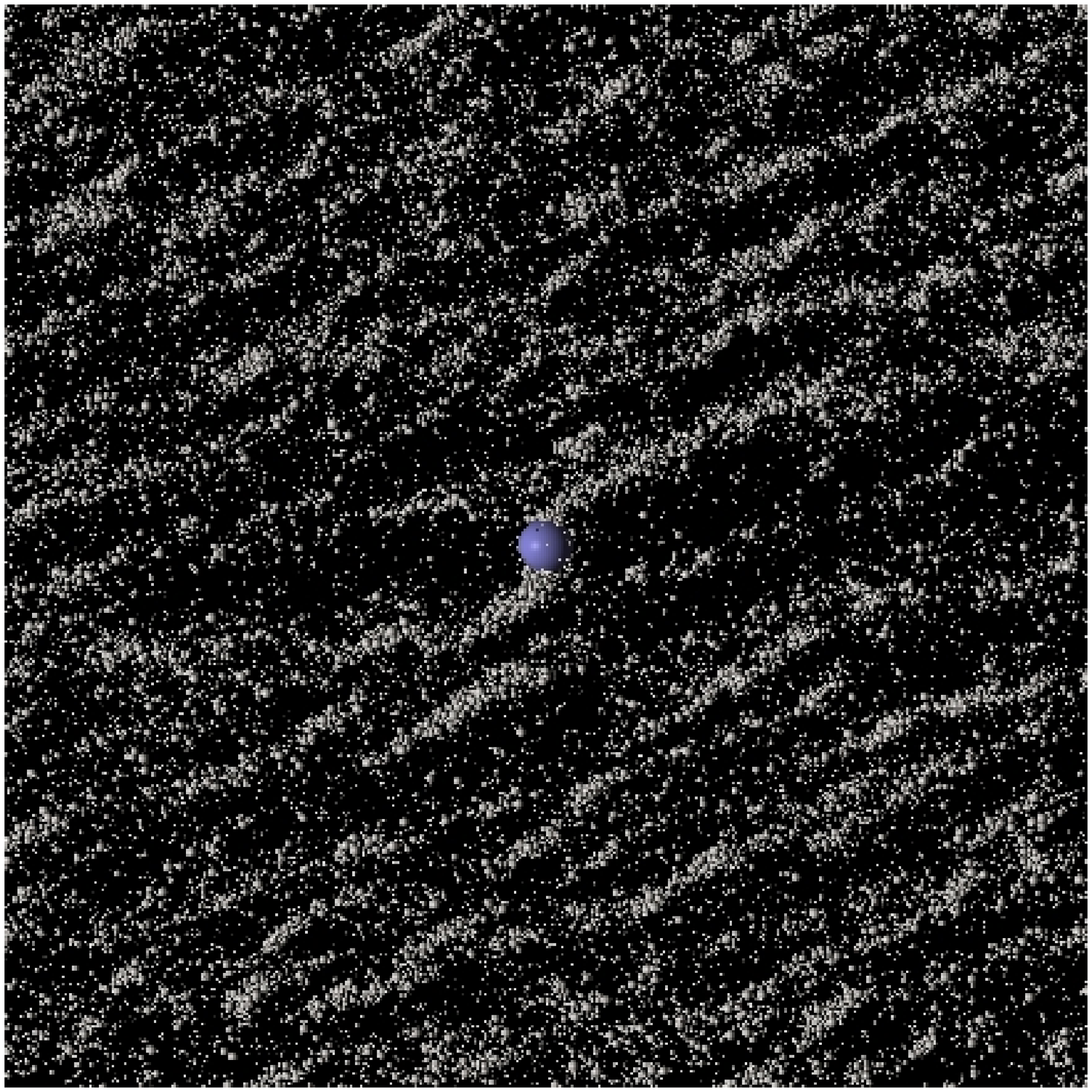}
}
\subfigure[$\Sigma=400\,\mathrm{kg/m^2}$, $r_2=50\,\mathrm{m}$]{
\includegraphics[angle=90,width=0.999\columnwidth]{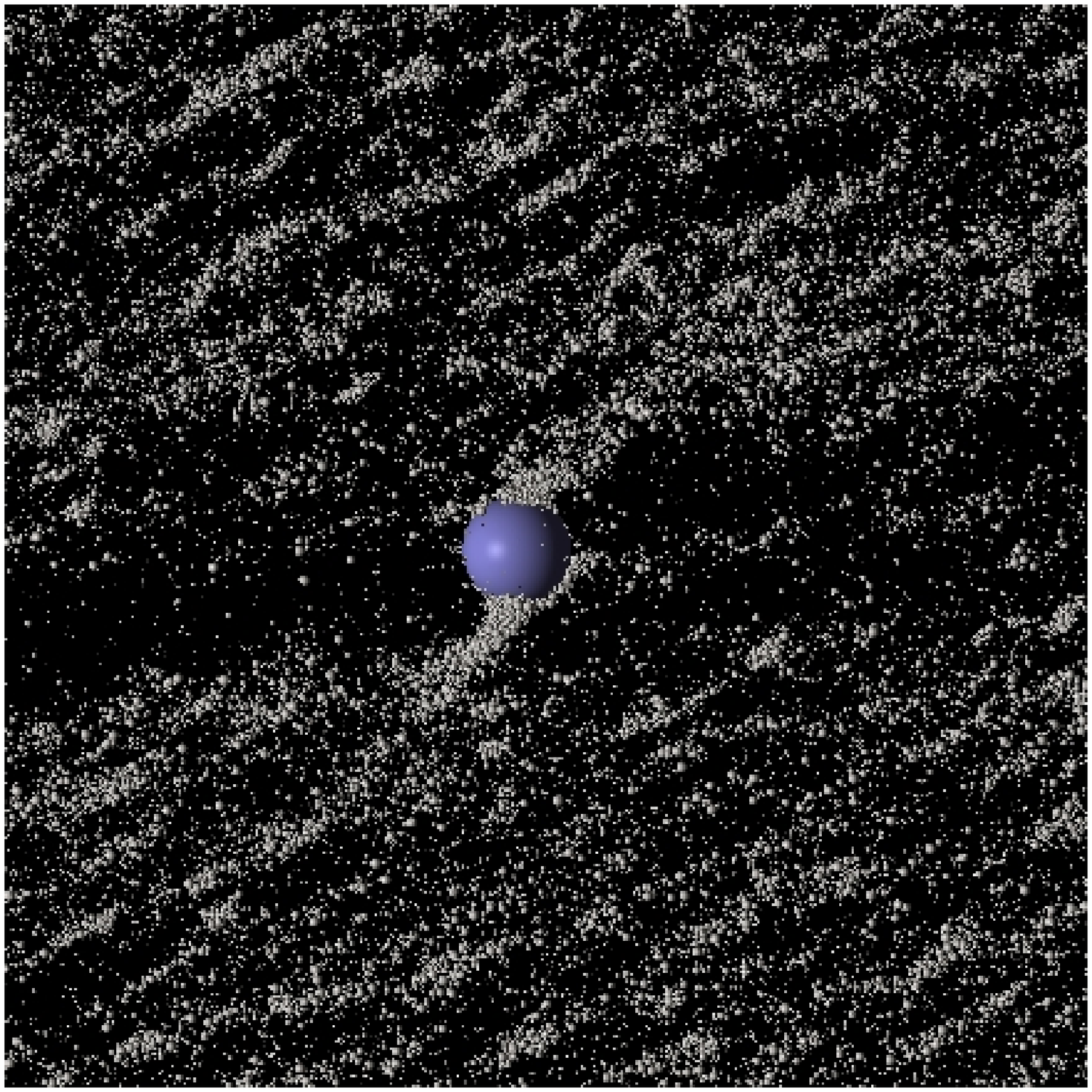}
}
\caption{Snapshots of the particle distribution and the moonlet (blue) for different surface densities after 25 orbits.
The wake created by the moonlet is much longer than the size of the box and therefore hardly visible in these images. 
\label{fig:snap}}
\end{figure*}


Throughout this paper, we use a distribution of particle sizes, $r_1$, ranging from $1$m to $5$m with a slope of $q=-3$.
Thus $dN/dr_1 \propto r_1^{-q}.$ 
The density of both the ring particles and the moonlet is taken to be $0.4 \,\mathrm{g/cm}^2$. 
This is at the lower end of what has been assumed reasonable for Saturn's A ring \citep{LewisStewart2009}. The moonlet radius is taken to be either $50\,\mathrm{m}$
or $25\,\mathrm{m}.$
We found that using a larger ring particle and moonlet density only leads to more particles being bound to the moonlet. 
This effectively increases the mass of the moonlet (or equivalently its Hill radius) and can therefore
be easily scaled to the formalism presented here. 
Simulating a gravitational aggregate of this kind
is computationally very expensive, as many more collisions have to be resolved each time-step. 
\clearpage
\begin{figure*}[t]
\centering
\includegraphics[angle=270,width=0.95\textwidth]{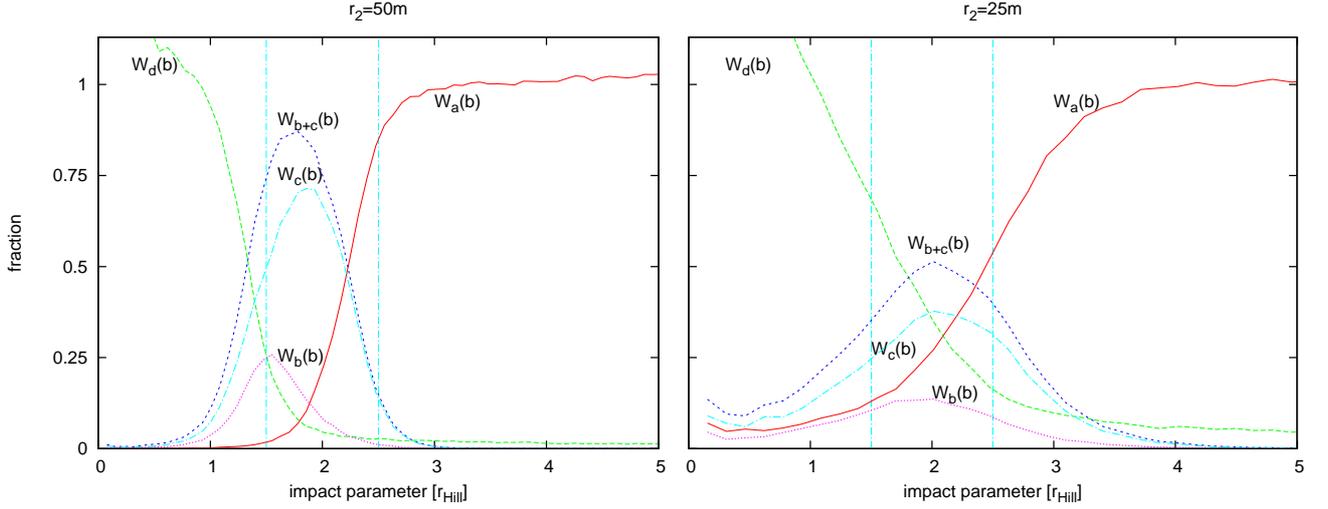}
\caption{Impact bands for simulation \texttt{EQ20050} (left) and \texttt{EQ20025} (right). 
The vertical lines indicate sharp cutoffs at $1.5r_H$ and $2.5r_H$.
\label{fig:bands_50m}}
\end{figure*}

The initial velocity dispersion of the particles is set to $1\,\mathrm{mm/s}$ for the $x$ and $y$ components and $0.4\,\mathrm{mm/s}$ for the $z$ component. 
The moonlet is placed at a semi-major axis of $a=130000\,\mathrm{km}$, corresponding to an orbital period of $P=13.3\,\mathrm{hours}$.
Initially, the moonlet is placed in the centre of the shearing box on a circular orbit. 

We run simulations with a variety of moonlet sizes, surface densities and box sizes. 
The dimensions of each box, as viewed in the $(x,y)$ plane, are specified in Table \ref{tab:sim}. 
Because of the small dispersion velocities, the vertical motion is automatically strongly confined to the mid-plane.
After a few orbits, the simulations reach an equilibrium state in which the velocity dispersion of particles does not change anymore.

We ran several tests to ensure that our results are converged. 
Simulation \texttt{EQ40050DT} uses a ten times larger time step than simulation \texttt{EQ40050}. 
Simulations \texttt{EQ5050DTW}, \texttt{EQ20050DTW} and \texttt{EQ40050DTW} use a box that is twice the size of that used in simulation \texttt{EQ40050}. 
Therefore, four times more particles have been used. 
No differences in the equilibrium state of the ring particles, nor in the equilibrium state of the moonlet have been observed in any of those test cases.

A snapshot of the particle distribution in simulation \texttt{EQ40050DTW} is shown in Fig. \ref{fig:snapdtw}.
Snapshots of simulations \texttt{EQ20025}, \texttt{EQ20050}, \texttt{EQ40025} and \texttt{EQ40050}, which use 
the smaller box size, are shown in Fig. \ref{fig:snap}.
In all these cases, the length of the wake that is created by the moonlet is much longer than the box size. 
Nevertheless, the pseudo shear periodic boundary conditions allow us to simulate the evolution of the moonlet accurately.

\section{Results} \label{sec:results}

\subsection{Impact bands}

The impact band window functions 
$W_a(b)$, $W_{b+c}(b)$ and $W_{d}(b)$ are necessary to estimate the damping timescales and the strength of the excitation.
To measure those, each particle that enters the box is labelled with it's impact parameter. 
The possible outcomes are plotted in Fig. \ref{fig:trajectory}: (a) being a circulating particle which leaves the box on the opposite side, (b) representing a collision with other ring particles close to the moonlet where the maximum distance 
from the centre of the moonlet 
has been taken to be twice the moonlet radius, (c) being a direct collision with the moonlet and finally (d) showing a horseshoe orbit in which the particle leaves the box on the same side that it entered. 

The impact band (b) is considered in addition to the impact band (c) because some ring particles will collide with other ring particles that are (temporarily) bound to the moonlet. Thus, these collisions take part in the transfer the energy and momentum to the moonlet. Actually, if the moonlet is simply a rubble pile of ring particles as suggested by \cite{Porco2007}, then there might be no solid moonlet core and all collisions are in the impact band (b). 

Fig. \ref{fig:bands_50m} shows the impact bands, normalised to the Hill radius of the moonlet $r_{\mathrm{H}}$ for simulations \texttt{EQ20050} and \texttt{EQ20025}\footnote{$\sum_{i=a,b,c,d}W_i\gtrsim1$ because all particles are shifted from time to time when the moonlet is to far away from the origin (see Sect. \ref{sec:methods}). In this process, particles with the same initial impact parameter might become associated with the more than one impact band.}.
For comparison, we also plot sharp cutoffs, at $1.5r_H$ and $2.5r_H$. 
Our results show no sharp discontinuity because of the velocity dispersion in the ring particles which is $\sim 5\,\mathrm{mm/s}$. 
This corresponds to an epicyclic motion of $\sim40\,\mathrm{m}$ or equivalently $\sim0.8\,r_H$ and $\sim1.6\,r_H$ (for $r_2=50\mathrm{m}$ and $r_2=25\mathrm{m}$, respectively), which explains the cutoff width of approximately one Hill radii. 
The impact bands are almost independent of the surface density and depend only on the Hill radii and the mean epicyclic amplitude of the ring particles, or, more precisely, the ratio thereof.

\subsection{Eccentricity damping timescale}
To measure the eccentricity damping timescale, we first let the ring particles and the moonlet reach an equilibrium and integrate them for 200 orbits. 
We then change the velocity of the moonlet and the ring particles within $2r_{\mathrm{H}}$. 
The new velocity corresponds to an eccentricity of $6\cdot 10^{-7}$, which is well above the equilibrium value. 
We then measure the decay timescale $\tau_e$ by fitting a function of the form 
\begin{eqnarray}
e_d(t)=\langle e_{eq}\rangle+\left(6\cdot10^{-7}-\langle e_{eq}\rangle\right) \; e^{-t/\tau_e}.
\end{eqnarray}
The results are given in Table \ref{tab:simresults1}. 
They are in good agreement with the estimated damping timescales from Sect. \ref{sec:edampcollision} and \ref{sec:edamphorseshoe}, showing clearly that the most important damping process, in every simulation considered here, is indeed through collisional damping as predicted by comparing Eq. \ref{eq:taucol} with Eqs. \ref{eq:damphorse} and \ref{Goldcirc2}.

\begin{table}[t]
\begin{center}
\begin{tabular}{l|r|rrr}
Name 		&\multicolumn{1}{c|}{Numerical results}	
		&\multicolumn{2}{c}{Analytic results }\\	
		
	& 	
	& collisions 
	& horseshoe 
		 	\\\hline\hline
\texttt{EQ5025}		
	& 18.9	& 18.4 & 1712 	\\	
\texttt{EQ5050}		
	& 27.6 	& 33.8 & 3424	\\
\texttt{EQ20025}	
	& 9.0	& 3.9 & 428	\\
\texttt{EQ20050}	
	& 12.9	& 8.5 & 856	\\
\texttt{EQ40025}	
	& 2.8	& 2.1 & 214	\\
\texttt{EQ40050}	
	& 5.5	& 4.4 & 428
\end{tabular}
\caption{Eccentricity damping timescale $\tau_e$ of the moonlet in units of the orbital period. The second column lists the simulation results. The third and fourth column list the analytic estimates of collisional and horseshoe damping timescales, respectively. \label{tab:simresults1}}
\end{center}
\end{table}

\subsection{Mean moonlet eccentricity}
The mean eccentricity of the moonlet
is measured in all simulations after several orbits when the ring particles and the moonlet have reached an equilibrium state. 
To compare this value with the estimates from Sect. \ref{sec:analytic} we set Eq. \ref{equile} equal to zero and use the analytic damping timescale listed in Table \ref{tab:simresults1}.
The analytic estimates of the equilibrium eccentricity are calculated for each excitation mechanism separately to disentangle their effects. 
They are listed in the third, fourth and fifth column in Table \ref{tab:simresults2}. 
The sixth column lists the analytic estimate for the mean eccentricity using the sum of all excitation mechanisms. 

For all simulations, the estimates are correct within a factor of about 2. 
For low surface densities, the excitation is dominated by individual particle collisions.
For larger surface densities, it is dominated by the excitation due to circulating self gravitating clumps (gravitational wakes). 
The estimates and their trends are surprisingly accurate, as we have ignored several effects (see below).

\subsection{Random walk in semi-major axis}
The random walk of the semi-major axis $a$ (or equivalently the centre of epicyclic motion ${\cal A}$ in the shearing sheet) of the moonlet in the numerical simulations are measured and compared to the analytic estimates presented in Sect. \ref{sec:randa}.
We ran one simulation per parameter set. 
To get a statistically meaningful expression for the average random walk after a given time ${\cal A}(\Delta t)$, we average all pairs of ${\cal A}(t)$ and ${\cal A}(t')$ for which $t-t'=\Delta t$. In other words, we assume the system satisfies the Ergodic hypothesis.

${\cal A}(\Delta t)$ then grows like $\sqrt{\Delta t}$ and we can fit a simple square root function.
This allows us to accurately measure the average growth in ${\cal A}$ after $\Delta t=100$ orbits by running one simulation for a long time and averaging over time, rather than running many simulations and performing an ensemble average.
The measured values are given in the second column of Table \ref{tab:simresults3}.

The values that correspond to the analytic expressions in Eqs. \ref{eq:randwalka:horseshoe}, \ref{eq:randwalka:collisions}, \ref{eq:randwalka:circulatingparticles} and \ref{eq:randwalka:circulatingclumps} are listed in columns three, four, five and six, respectively.

For low surface densities, the evolution of the random walk is dominated by collisions and the effect of particles on horseshoe orbits. 
For large surface densities, the main effect comes from the stochastic gravitational force due to circulating clumps (gravitational wakes).


\begin{table*}[t]
\begin{center}
\begin{tabular}{l|r|rrr|r}
Name &\multicolumn{1}{c|}{Numerical results}	
		&\multicolumn{4}{c}{Analytic results}\\	
	&
	& collisions
	& circulating particles
	& \multicolumn{1}{c|}{circulating clumps}
	& \multicolumn{1}{c}{total}
		 	\\\hline\hline
\texttt{EQ5025}	&	
	$6.1\cdot10^{-8}$	&$4.5\cdot10^{-8}$ &$1.9\cdot10^{-8}$ &$0.3\cdot10^{-8}$ &$4.9\cdot10^{-8}$ \\	
\texttt{EQ5050}	& 	
	$4.3\cdot10^{-8}$ 	&$1.6\cdot10^{-8}$ &$1.4\cdot10^{-8}$ &$0.2\cdot10^{-8}$ &$2.1\cdot10^{-8}$ \\
\texttt{EQ20025}&	
	$6.4\cdot10^{-8}$	&$4.6\cdot10^{-8}$ &$1.7\cdot10^{-8}$ &$2.0\cdot10^{-8}$ &$5.3\cdot10^{-8}$ \\
\texttt{EQ20050}&	
	$4.9\cdot10^{-8}$ 	&$1.6\cdot10^{-8}$ &$1.4\cdot10^{-8}$ &$1.6\cdot10^{-8}$ &$2.7\cdot10^{-8}$ \\
\texttt{EQ40025}&	
	$11.3\cdot10^{-8}$ 	&$4.6\cdot10^{-8}$ &$2.5\cdot10^{-8}$ &$8.4\cdot10^{-8}$ &$9.9\cdot10^{-8}$ \\
\texttt{EQ40050}&	
	$7.2\cdot10^{-8}$ 	&$1.6\cdot10^{-8}$ &$1.5\cdot10^{-8}$ &$4.9\cdot10^{-8}$ &$5.4\cdot10^{-8}$
\end{tabular}
\caption{Equilibrium eccentricity $\langle e_{\mathrm{eq}}\rangle$ of the moonlet. 
The second column lists the simulation results. 
The third, fourth and fifth column list the analytic estimates of the equilibrium eccentricity assuming a single excitation mechanism. 
The last column lists the analytic estimate of the equilibrium eccentricity summing over all excitation mechanisms. \label{tab:simresults2}}
\end{center}
\end{table*}

\begin{table*}[t]
\begin{center}
\begin{tabular}{l|r|rrrr|r}
Name 		&\multicolumn{1}{c|}{Numerical results}	
		&\multicolumn{4}{c}{Analytic results }\\	
		
	& 	
	& horseshoe
	& collisions 
	& circulating particles 
	& circulating clumps
	& total 
		 	\\\hline\hline
\texttt{EQ5025}		
	&22.7m 		& 9.0m	& 10.5m 	& 17.6m 	& 0.3m		& 22.4m 	\\	
\texttt{EQ5050}		
	&12.6m 		& 4.5m	& 4.8m 		& 4.4m 		& 0.1m		& 7.9m	 	\\
\texttt{EQ20025}	
	&38.1m 		& 18.1m	& 25.1m 	& 17.6m 	& 15.7m		& 38.9m		\\
\texttt{EQ20050}	
	&22.7m 		& 13.6m	& 9.6m 		& 4.4m 		& 4.8m		& 14.7m 	\\
\texttt{EQ40025}	
	&78.1m 		& 25.6m	& 36.5m 	& 17.6m	 	& 86.6m		& 100.7m	\\
\texttt{EQ40050}	
	&45.6m 		& 12.8m	& 13.4m 	& 4.4m 		& 31.4m		& 36.7m
\end{tabular}
\caption{Change in semi-major axis of the moonlet after 100 orbits, ${\cal A}(\Delta t=100\,\mathrm{orbits})$. 
The second column lists the simulation results. 
The third till sixth columns list the analytic estimates of the change in semi-major axis assuming a single excitation mechanism.
The last column lists the total estimated change in semi-major axis summing over all excitation mechanisms. \label{tab:simresults3}}
\end{center}
\end{table*}


\subsection{Long term evolution and observability}
\begin{figure}[t]
\centering
	\includegraphics[angle=270,width=0.95\columnwidth]{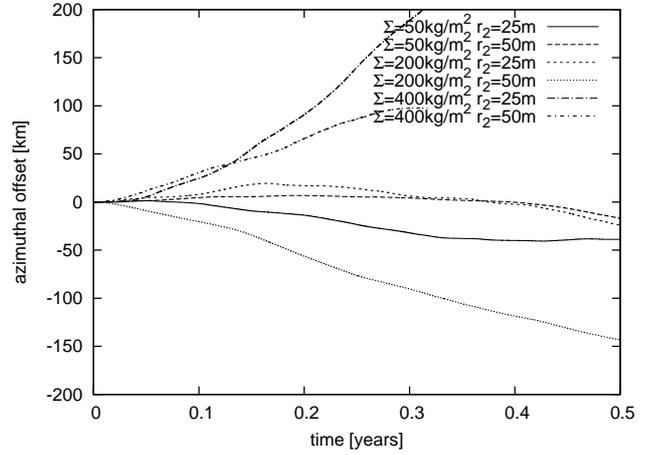}
\caption{Offset in the azimuthal distance due to the random walk in the semi-major axis $a$ measured in simulations \texttt{EQ5025}, \texttt{EQ5050}, \texttt{EQ20025}, \texttt{EQ20050}, \texttt{EQ40025} and \texttt{EQ40050}. Individual moonlets may show linear, constant or oscillatory growth. On average, the azimuthal offset grows like $t^{3/2}$.\label{fig:azimuth}}
\end{figure}
The change in semi-major axis is relatively small, a few tens of meters after 100 orbits ($=50$ days). 
The change in longitude that corresponds to this is, however, is much larger. 
We therefore extend the discussion of \cite{ReinPapaloizou2009}, who derive the time evolution of orbital parameters undergoing a random walk, to the time evolution of the mean longitude.
This is of special interest in the current situation, as in observations of Saturn's rings, it is much easier to measure the shift in the longitude of an embedded moonlet than the change in its semi-major axis.

The time derivative of the mean motion is given by
\begin{eqnarray}
\dot n = - \frac32 \sqrt{\frac{G M}{a^5} }\dot a = - \frac{3\; F_\theta }{a},
\end{eqnarray}
where $F_\theta$ is the time dependent stochastic force in the $\theta$ (or in shearing sheet coordinates, $y$) direction and we have assumed a nearly circular orbit, $e\ll 1$. The mean longitude is thus given by the double integral
\begin{eqnarray}
\lambda(t) 
	&=& \int_0^t \left(n_0+\Delta n(t')\right)\;dt'\\
	&=& n_0 t - \int_0^t \int_0^{t'} 		\frac{3F_\theta(t'')}a		\;dt''dt'.
\end{eqnarray}
The root mean square of the difference compared to the unperturbed orbit ($n_0$) is
\begin{eqnarray}
\left(\Delta \lambda\right)^2 &=&
\left<\left(\lambda(t)-nt\right)^2 \right>\\
	&=& \iiiint\limits_0^{t,\,t',\,t,\,t'''} \frac{9F_\theta(t'')F_\theta(t'''')}{a^2}		\;dt''''dt'''dt''dt'\\
	&=& \frac{9\left< F_\theta^2 \right>}{a^2} \iiiint\limits_0^{t,\,t',\,t,\,t'''}	 g(|t''-t''''|) 			\;dt''''dt'''dt''dt' \label{eq:randwalk:notdoable}\\
	&=& \frac{9\left< F_\theta^2 \right>}{a^2} \left(
	-2\tau^4+
	\left( 2\tau^3t+2\tau^4+\tau^2t^2 \right ) e^{-t/\tau}
	+\frac 13 \tau t^3
	\right) \label{eq:randwalk:allterms}
\end{eqnarray}
where $g(t)$ is the auto-correlation function of $F_\theta$, which has been assumed to be exponentially decaying with an auto-correlation time $\tau,$
thus $g(t)=\exp(-|t|/\tau).$
In the limit $t\gg\tau$, the above equation simplifies to 
\begin{eqnarray}
\left(\Delta \lambda\right)^2 &=&
	\frac{3\left< F_\theta^2 \right> \tau }{a^2} \;t^3
	\;= \left(\Delta n\right)^2 \frac{t^2}6\label{eq:growthlambda}.
\end{eqnarray} 
On shorter timescale, the other terms in Eq. \ref{eq:randwalk:allterms} are important, leading to linear and oscillatory behaviour.
For uncorrelated noise (e.g. collisions) Eq. \ref{eq:growthlambda} is replaced by 
\begin{eqnarray}
\left(\Delta \lambda\right)^2 &=&
	\frac{3\left< \Delta v \right>^2 }{a^2} \;t^3/\tau,
\end{eqnarray} 
where $\left<\Delta v \right>$ is the average velocity change per impulsive event and $\tau$ is taken to be the average time between consecutive events.

Instead of a stochastic migration, let us also assume a laminar migration with constant migration rate $\tau_a$, defined by
\begin{eqnarray}
\tau_a&=&\frac{a}{\dot a}=-\frac32 \frac{n}{\dot n}.
\end{eqnarray}
As above, we can calculate the longitude as a function of time as the following double integral
\begin{eqnarray}
\lambda(t) 
	&=& \int_0^t \left(n_0+\Delta n(t')\right)\;dt'\\
	&=& n_0 t - \int_0^t \int_0^{t'} 		\frac32 \frac{n}{\tau_a}		\;dt''dt'
	= n_0 t - \frac34 \frac{n}{\tau_a}\;t^2		
\end{eqnarray}
and thus
\begin{eqnarray}
\left(\Delta \lambda\right)^2 &=&\frac{9}{16} \frac{n^2}{\tau_a^2}\;t^4 = \left( \Delta n\right)^2 \frac{t^2}4.\label{eq:growthlambdalaminar}
\end{eqnarray}
Note that Eq. \ref{eq:growthlambdalaminar} is an exact solution, whereas Eq. \ref{eq:growthlambda} is a statistical quantity, describing the root mean square value in an ensemble average. 
In the stochastic and laminar case, $\left(\Delta\lambda\right)^2$ grows like $\sim t^3$ and $\sim t^4$, respectively. 
This behaviour allows to discriminate between a stochastic and laminar migration by observing $\Delta \lambda$ over an extended period of time. 

Results from individual simulations (i.e. not an ensemble average) are plotted in Fig. \ref{fig:azimuth}. Here, $\Delta \lambda$ is expressed in terms of the azimuthal offset relative to a Keplerian orbit. 
One can see that for an individual moonlet, the shift in azimuth can appear to be linear, constant or oscillating on short timescales (see curve for simulation \texttt{EQ20025}). 
This is partly because of the lower order terms in Eq. \ref{eq:randwalk:allterms}. 
Though, on average, the azimuthal offset grows very rapidly, as $\sim t^{3/2}$ (see Eq. \ref{eq:growthlambda}) for $t\gg\tau$.

\section{Conclusions}\label{sec:conclusions}
In this paper, we have discussed the dynamical response
of an embedded moonlet in Saturn's rings to interactions with ring particles both analytically and by the use of realistic three dimensional many particle simulations. 
Both the moonlet and the ring particle density were taken to be $0.4 \,\mathrm{g/cm}^2.$ Moonlets of radius $25\,\mathrm{m}$ and $50\,\mathrm{m}$ were considered. 
Particle sizes ranging between $1\,\mathrm{m}$ and $5\,\mathrm{m}$ were adopted.

We estimated the eccentricity damping timescale of the moonlet due to collisions with ring particles and due to the response of ring particles to gravitational perturbations by the moonlet analytically. 
We found the effects due to the response of particles on horseshoe and circulating orbits are negligible.
On the other hand the stochastic impulses applied to the moonlet by circulating particles were found to cause the square of the eccentricity to grow linearly with time as did collisions with particles with non zero velocity dispersion.
A balance between excitation and damping processes then leads to an equilibrium moonlet eccentricity. 

We also estimated the magnitude of the random walk in the semi-major axis of the moonlet induced by collisions with individual ring particles and the gravitational interaction with particles and gravitational wakes. 
There is no tendency for the semi-major axis to relax to any particular value, so that there are no damping terms.
The deviation of the semi-major axis from its value at time $t=0$ grows on average as $\sqrt{t}$ for large $t.$

From our simulations we find that the evolution of the eccentricity is indeed dominated by collisions with ring particles. 
For large surface densities (more than $200\mathrm{kg/m^2}$) the effect of gravitational wakes also becomes important, leading to an increase in the mean steady state eccentricity of the moonlet. 
When the particle velocity dispersion is large compared to $\Omega r_H,$ we obtain approximate energy equipartition between the moonlet and ring particles as far as epicyclic motion is concerned.

Similarly, the random walk in the semi-major axis was found to be dominated by collisions for low surface densities and by gravitational wakes for large surface densities. 
In addition we have shown that on average, the difference in longitude $\Delta \lambda$ of a stochastically forced moonlet grows with time like $t^{3/2}$ for large~$t.$ 

The distance travelled within 100 orbits (50 days at a distance of 130000km) is, depending on the precise parameters, of the order of 10-100m. 
This translates to a shift in longitude of several hundred kilometres. 
The shift in $\lambda$ is not necessarily monotonic on short timescales (see Fig. \ref{fig:azimuth}).
We expect that such a shift should be easily detectable by the Cassini spacecraft.
And indeed, \cite{Tiscareno2010} report such an observation, although the data is not publicly available yet, at the time when this paper was submitted.

There are several effects that have not been included in this study.
The analytic discussions in Sec. \ref{dampe}, Sec. \ref{sec:excitee} and Sec. \ref{sec:randa} do not model the motion of particles correctly when their trajectories take them very close to the moonlet. 
Material that is temporarily or permanently bound to the moonlet is ignored. 
The density of both the ring particles and the moonlet are at the lower end of what is assumed to be reasonable \citep{LewisStewart2009}. 
This has the advantage that for our computational setup only a few particles (with a total mass of less than 10\% of the moonlet) are bound to the moonlet at any given time.
In a future study, we will extend the discussion presented in this paper to a larger variety of particle sizes, moonlet sizes and densities.

\begin{acknowledgements}
We thank Aur\'{e}lien Crida for reading an early draft of this paper. 
Hanno Rein was supported by an Isaac Newton Studentship and St John's College Cambridge. 
Simulations were performed on the computational facilities of the Astrophysical Fluid Dynamics Group and on Darwin, the Cambridge University HPC facility.
\end{acknowledgements}

\begin{appendix}
\section{Response calculation of particles on horseshoe orbits}\label{app:response}
\subsection{Interaction potential}
Due to some finite eccentricity, the moonlet undergoes a small oscillation about the origin.
Its Cartesian coordinates then become $(X,Y)$, with these being considered small in magnitude.
The components of the equation of motion for a ring particle with Cartesian coordinates $(x,y)\equiv {\bf r}_1$ are
\begin{eqnarray}
\frac{d^2 x}{dt^2}-2\Omega\frac{d y}{dt} &=&3\Omega^2x - \frac{1}{m_1}\frac{\partial \Psi_{1,2}}{\partial x}\hspace{3mm} {\rm and}\\
\frac{d^2 y}{dt^2}+2\Omega\frac{d x}{dt} &=&-\frac{1}{m_1} \frac{\partial \Psi_{1,2}}{\partial y},\label{equmot}
\end{eqnarray}
where the interaction gravitational potential due to the moonlet is
\begin{eqnarray}
\Psi_{1,2} = -\frac{Gm_1m_2}{(r^2+R^2-2rR\cos(\phi -\Phi))^{1/2}}
\end{eqnarray} 
with the cylindrical coordinates of the particle and moonlet being $(r, \phi)$ and $(R,\Phi)$ respectively.
This may be expanded correct to first order in $R/r$ in the form
\begin{eqnarray}
\Psi_{1,2} = -\frac{Gm_1m_2}{r} -\frac{Gm_1m_2R\cos(\phi -\Phi)}{r^2}.\label{pertpot}
\end{eqnarray}
The moonlet undergoes small amplitude epicyclic oscillations
such that $X={\cal E}_2\cos(\Omega t +\epsilon), Y=-2{\cal E}_2\sin(\Omega t +\epsilon)$
where $e$ is its small eccentricity and $\epsilon$ is an arbitrary phase.
Then $\Psi_{1,2}$ may be written as
\begin{eqnarray}
\Psi_{1,2} = -\frac{Gm_1m_2}{r} - \Psi'_{1,2},
\end{eqnarray}
where
\begin{eqnarray}
\Psi'_{1,2} = - \frac{Gm_1m_2r\,{\cal E}_2(x\cos(\Omega t +\epsilon) -2y\sin(\Omega t +\epsilon))}{r^3}
\label{pertpot1}\end{eqnarray}
gives the part of the lowest order interaction potential associated with 
the eccentricity of the moonlet.

We here view the interaction of a ring particle with the moonlet as involving two components.
The first, due to the first term on the right hand side of Eq. \ref{pertpot} operates when ${\cal E}_2=0$
and results in standard horseshoe orbits for ring particles induced by a moonlet in circular orbit.
The second term, Eq. \ref{pertpot1}, perturbs this motion when ${\cal E}_2$ is small.
We now consider the response of a ring particle undergoing horseshoe motion to this perturbation.
In doing so we make the approximation that the variation of the leading order potential Eq. \ref{pertpot}
due to the induced ring particle perturbations may be neglected. This is justified by the fact
that the response induces epicyclic oscillations of the particle which are governed
by the dominant central potential.

\subsection{Response calculation}
Setting $x\rightarrow x + \xi_x, y\rightarrow y + \xi_y,$ where \mbox{\boldmath${\xi}$}
is the small response displacement induced by Eq. \ref{pertpot1} and linearising Eqs. \ref{equmot},
we obtain the following equations for the components of \mbox{\boldmath${\xi}$}
\begin{eqnarray}
\frac{d^2\xi_x}{dt^2}-2\Omega\frac{d \xi_y}{dt} &=&3\Omega^2\xi_x - \frac{1}{m_1}\frac{\partial \Psi'_{1,2}}{\partial x}\hspace{3mm} {\rm and}\\
\frac{d^2 \xi_y}{dt^2}+2\Omega\frac{d \xi_x}{dt} &=&-\frac{1}{m_1} \frac{\partial \Psi'_{1,2}}{\partial y},\label{equpert}.
\end{eqnarray}
From these we find a single equation for $\xi_x$ in the form
\begin{eqnarray}
\frac{d^2\xi_x}{dt^2}+\Omega^2\xi_x &=& - \frac{1}{m_1}\frac{\partial \Psi'_{1,2}}{\partial x}-
\int \frac{1}{m_1}\frac{\partial \Psi'_{1,2}}{\partial y}dt = F
\label{equpert1}.
\end{eqnarray}
When performing the time integral on the right hand side of the above, as we are concerned with a potentially resonant epicyclic response,
we retain only the oscillating part. 

The solution to Eq. \ref{equpert1} which is such that \mbox{\boldmath${\xi}$} vanishes in the distant past $(t=-\infty)$ 
when the particle is far from the moonlet may be written as
\begin{eqnarray}
\xi_x=\alpha\cos(\Omega t)+ \beta\sin(\Omega t),
\end{eqnarray}
where
\begin{eqnarray}
\alpha & = & -\int_{-\infty}^t\frac{F\sin(\Omega t)}{\Omega}dt \hspace{3mm} {\rm and}\\
\beta & = & \int_{-\infty}^t\frac{F\cos(\Omega t)}{\Omega}dt.
\end{eqnarray}
After the particle has had its closest approach to the moonlet and moves to a large distance from it
it will have an epicyclic oscillation with amplitude and phase determined by
\begin{eqnarray}
\alpha_{\infty} & = & -\int_{-\infty}^{\infty}\frac{F\sin(\Omega t)}{\Omega}dt \hspace{3mm} {\rm and} \\
\beta_{\infty} & = & \int_{-\infty}^{\infty} \frac{F\cos(\Omega t)}{\Omega}dt.\label{epiamp}
\end{eqnarray}
In evaluating the above we note that, although $F$ vanishes when the particle
is distant from the moonlet at large $|t|,$ it also oscillates with the epicyclic angular frequency $\Omega$
which results in a definite non zero contribution. This is the action of the co-orbital resonance.
It is amplified by the fact that the encounter of the particle with the moonlet in general
occurs on a horseshoe libration time scale which is much longer than $\Omega^{-1}.$
We now consider this unperturbed motion of the ring particles.
\subsection{Unperturbed horseshoe motion}
The equations governing the unperturbed horseshoe motion are Eqs. \ref{equmot} 
with 
\begin{eqnarray}
\Psi_{1,2} = \Psi^0_{1,2}= -\frac{Gm_1m_2}{r}\label{horsepot}.
\end{eqnarray}
We assume that this motion is such that $x$ varies on a time scale much longer than $\Omega^{-1}$ so that we may approximate
the first of these equations as
$x=-2/(3\Omega)(dy/dt).$ Consistent with this we also neglect $x$ in comparison to $y$ in Eq. \ref{horsepot}.
From the second equation we than find that
\begin{eqnarray}
\frac{d^2 y}{dt^2} &=&\frac{3}{m_1} \frac{\partial \Psi^0_{1,2}}{\partial y},\label{equhorse}
\end{eqnarray}
which has a first integral that may be written
\begin{eqnarray}
\left(\frac{d y}{dt}\right)^2 = -\frac{6Gm_2}{|y|} +\frac{9\Omega^2b^2}{4},\label{HSHOE}
\end{eqnarray}
where as before $b$ is the impact parameter, or the constant value of $|x|$ at large distances from the moonlet.
The value of $y$ for which the horseshoe turns is then given by
$y=y_0=24Gm_2/(9\Omega^2b^2).$
At this point we comment that we are free to choose the origin of time $t=0$
such that it coincides with the closest approach at which $y=y_0.$
Then in the approximation we have adopted, the horseshoe motion 
is such that $y$ is a symmetric function of $t.$
\subsection{Evaluation of the induced epicyclic amplitude}
Using Eqs. \ref{pertpot1}, \ref{equpert1} and \ref{HSHOE}
we may evaluate the epicyclic amplitudes from Eq. \ref{epiamp}.
In particular we find 
\begin{eqnarray}
F= - \frac{Gm_2\,{\cal E}_2\cos(\Omega t +\epsilon)}{r^3}\left(5 -\frac{(3x^2+12y^2)}{r^2}\right). 
\end{eqnarray}
When evaluating Eq. \ref{epiamp}, consistent with our assumption that the epicyclic oscillations
are fast, we average over an orbital period assuming that other quantities in the integrands
are fixed, and use Eq. \ref{HSHOE} to express the integrals with respect to $t$ as integrals with respect to $y.$
We then find $\alpha_{\infty}=-A_{`\infty}\sin\epsilon $ and $\beta_{\infty}=-A_{`\infty}\cos\epsilon$ where
\begin{eqnarray}
A_{\infty}&=& \int^{\infty}_{y_0}\frac{\sqrt{6Gm_2y_0}\;{\cal E}_2}{6\Omega r^3\sqrt{1-y_0/y}}\nonumber
\\ &&\quad\cdot\left(5 -\frac{\left(\frac{8Gm_2}{\Omega^2}
\left(\frac{1}{y}-\frac{1}{y_0}\right)+12y^2\right)}{r^2}\right)dy,
\end{eqnarray}
with $r^2 = 8Gm_2/(3\Omega^2y_0)(1-y_0/y)+y^2.$
From this we may write
\begin{eqnarray}
|A_{\infty}|={\cal E}_{1f} =\frac{Gm_2\,{\cal E}_2}{y_0^3\Omega^2}|\cal{I}|,\label{inducede}
\end{eqnarray}
where ${\cal E}_{1f}$ is the final epicyclic motion of the ring particle and the dimensionless integral $\cal{I}$
is given by
\begin{eqnarray}
{\cal{I}} &=& \sqrt{\frac{\Omega^2 y^3_0}{6Gm_2}}\int^{\infty}_{y_0}\frac{y_0^2r^{-3}}{\sqrt{1-y_0/y}}\nonumber\\
&&\quad\cdot\left(5 -\frac{\left(\frac{8Gm_2}{\Omega^2}
\left(\frac{1}{y}-\frac{1}{y_0}\right)+12y^2\right)}{r^2}\right)dy.
\end{eqnarray}
We remark that the dimensionless quantity $\eta = 8Gm_2/(3\Omega^2y_0^3)=8(r_H/y_0)^3,$ with $r_H$ being the Hill radius
of the moonlet. It is related to the impact parameter $b$ by $\eta =2^{-6}(b/r_H)^6.$
Thus for an impact parameter amounting to a few Hill radii, in a very approximate sense, $\eta$ is of order unity, $y_0$ is of order $r_H$ and
the induced eccentricity ${\cal E}_{1f}$ is of order ${\cal E}_2$.
\end{appendix}

\bibliography{full}
\bibliographystyle{plainnat}

\label{lastpage}
\end{document}

%% file: trajectory.pstex_t
\begin{picture}(0,0)%
\includegraphics{trajectory.pstex}%
\end{picture}%
\setlength{\unitlength}{3947sp}%
\begingroup\makeatletter\ifx\SetFigFont\undefined%
\gdef\SetFigFont#1#2#3#4#5{%
  \reset@font\fontsize{#1}{#2pt}%
  \fontfamily{#3}\fontseries{#4}\fontshape{#5}%
  \selectfont}%
\fi\endgroup%
\begin{picture}(11724,5724)(-611,-7198)
\put(-74,-6211){\makebox(0,0)[lb]{\smash{{\SetFigFont{12}{14.4}{\familydefault}{\mddefault}{\updefault}{\color[rgb]{0,0,0}$x$}%
}}}}
\put(-524,-6811){\makebox(0,0)[lb]{\smash{{\SetFigFont{12}{14.4}{\familydefault}{\mddefault}{\updefault}{\color[rgb]{0,0,0}$y$}%
}}}}
\put(751,-2011){\makebox(0,0)[lb]{\smash{{\SetFigFont{12}{14.4}{\familydefault}{\mddefault}{\updefault}{\color[rgb]{0,0,0}(a)}%
}}}}
\put(751,-2686){\makebox(0,0)[lb]{\smash{{\SetFigFont{12}{14.4}{\familydefault}{\mddefault}{\updefault}{\color[rgb]{0,0,0}(b)}%
}}}}
\put(751,-3361){\makebox(0,0)[lb]{\smash{{\SetFigFont{12}{14.4}{\familydefault}{\mddefault}{\updefault}{\color[rgb]{0,0,0}(c)}%
}}}}
\put(751,-4036){\makebox(0,0)[lb]{\smash{{\SetFigFont{12}{14.4}{\familydefault}{\mddefault}{\updefault}{\color[rgb]{0,0,0}(d)}%
}}}}
\put(4501,-2686){\makebox(0,0)[lb]{\smash{{\SetFigFont{12}{14.4}{\familydefault}{\mddefault}{\updefault}{\color[rgb]{0,0,0}Lagrange point L1}%
}}}}
\put(4051,-5386){\makebox(0,0)[lb]{\smash{{\SetFigFont{12}{14.4}{\familydefault}{\mddefault}{\updefault}{\color[rgb]{0,0,0}Roche lobe}%
}}}}
\put(751,-5536){\makebox(0,0)[lb]{\smash{{\SetFigFont{12}{14.4}{\familydefault}{\mddefault}{\updefault}{\color[rgb]{0,0,0}(e)}%
}}}}
\put(4501,-6136){\makebox(0,0)[lb]{\smash{{\SetFigFont{12}{14.4}{\familydefault}{\mddefault}{\updefault}{\color[rgb]{0,0,0}Lagrange point L2}%
}}}}
\put(301,-4186){\rotatebox{90.0}{\makebox(0,0)[lb]{\smash{{\SetFigFont{12}{14.4}{\familydefault}{\mddefault}{\updefault}{\color[rgb]{0,0,0}impact parameter $b$}%
}}}}}
\end{picture}%

%% file: boxes.pstex_t
\begin{picture}(0,0)%
\includegraphics{boxes.pstex}%
\end{picture}%
\setlength{\unitlength}{3276sp}%
\begingroup\makeatletter\ifx\SetFigFont\undefined%
\gdef\SetFigFont#1#2#3#4#5{%
  \reset@font\fontsize{#1}{#2pt}%
  \fontfamily{#3}\fontseries{#4}\fontshape{#5}%
  \selectfont}%
\fi\endgroup%
\begin{picture}(5919,5124)(2014,-6148)
\put(7201,-2761){\makebox(0,0)[lb]{\smash{{\SetFigFont{10}{12.0}{\familydefault}{\mddefault}{\updefault}{\color[rgb]{0,0,0}shear}%
}}}}
\put(2401,-5161){\makebox(0,0)[lb]{\smash{{\SetFigFont{10}{12.0}{\familydefault}{\mddefault}{\updefault}{\color[rgb]{0,0,0}shear}%
}}}}
\put(6826,-5386){\makebox(0,0)[lb]{\smash{{\SetFigFont{10}{12.0}{\familydefault}{\mddefault}{\updefault}{\color[rgb]{0,0,0}auxiliary boxes}%
}}}}
\put(7201,-4186){\makebox(0,0)[lb]{\smash{{\SetFigFont{10}{12.0}{\familydefault}{\mddefault}{\updefault}{\color[rgb]{0,0,0}y}%
}}}}
\put(5026,-6136){\makebox(0,0)[lb]{\smash{{\SetFigFont{10}{12.0}{\familydefault}{\mddefault}{\updefault}{\color[rgb]{0,0,0}x}%
}}}}
\put(4576,-3736){\makebox(0,0)[lb]{\smash{{\SetFigFont{10}{12.0}{\familydefault}{\mddefault}{\updefault}{\color[rgb]{0,0,0}main box}%
}}}}
\put(5026,-1486){\rotatebox{270.0}{\makebox(0,0)[lb]{\smash{{\SetFigFont{10}{12.0}{\familydefault}{\mddefault}{\updefault}{\color[rgb]{0,0,0}to planet}%
}}}}}
\end{picture}%

%% file: main.bbl
\begin{thebibliography}{18}
\expandafter\ifx\csname natexlab\endcsname\relax\def\natexlab#1{#1}\fi
\expandafter\ifx\csname url\endcsname\relax
  \def\url#1{{\tt #1}}\fi

\bibitem[{Barnes} and {Hut}(1986)]{Barnes1986}
J.~{Barnes} and P.~{Hut}.
\newblock {A Hierarchical O(NlogN) Force-Calculation Algorithm}.
\newblock {\em \nat}, 324:\penalty0 446--449, December 1986.

\bibitem[{Bridges} et~al.(1984){Bridges}, {Hatzes}, and {Lin}]{Bridges1984}
F.~G. {Bridges}, A.~{Hatzes}, and D.~N.~C. {Lin}.
\newblock {Structure, stability and evolution of Saturn's rings}.
\newblock {\em \nat}, 309:\penalty0 333--335, May 1984.

\bibitem[{Crida} et~al.(2010){Crida}, {Papaloizou}, {Rein}, {Charnoz}, and
  {Salmon}]{Crida2010}
A.~{Crida}, J.~{Papaloizou}, H.~{Rein}, S.~{Charnoz}, and J.~{Salmon}.
\newblock Migration of a moonlet in a ring of solid particles: Theory and
  application to saturn's propellers.
\newblock {\em AJ}, 2010.
\newblock Submitted.

\bibitem[{Daisaka} et~al.(2001){Daisaka}, {Tanaka}, and {Ida}]{Daisaka2001}
H.~{Daisaka}, H.~{Tanaka}, and S.~{Ida}.
\newblock {Viscosity in a Dense Planetary Ring with Self-Gravitating
  Particles}.
\newblock {\em Icarus}, 154:\penalty0 296--312, December 2001.

\bibitem[{Goldreich} and {Tremaine}(1980)]{GoldreichTremaine1980}
P.~{Goldreich} and S.~{Tremaine}.
\newblock {Disk-satellite interactions}.
\newblock {\em \apj}, 241:\penalty0 425--441, October 1980.

\bibitem[{Goldreich} and {Tremaine}(1982)]{GoldreichTremaine1982}
P.~{Goldreich} and S.~{Tremaine}.
\newblock {The dynamics of planetary rings}.
\newblock {\em \araa}, 20:\penalty0 249--283, 1982.

\bibitem[{Lewis} and {Stewart}(2009)]{LewisStewart2009}
M.~C. {Lewis} and G.~R. {Stewart}.
\newblock {Features around embedded moonlets in Saturn's rings: The role of
  self-gravity and particle size distributions}.
\newblock {\em Icarus}, 199:\penalty0 387--412, February 2009.

\bibitem[{Morishima} and {Salo}(2006)]{MorishimaSalo2006}
R.~{Morishima} and H.~{Salo}.
\newblock {Simulations of dense planetary rings. IV. Spinning self-gravitating
  particles with size distribution}.
\newblock {\em Icarus}, 181:\penalty0 272--291, March 2006.

\bibitem[{Porco} et~al.(2007){Porco}, {Thomas}, {Weiss}, and
  {Richardson}]{Porco2007}
C.~C. {Porco}, P.~C. {Thomas}, J.~W. {Weiss}, and D.~C. {Richardson}.
\newblock {Saturns Small Inner Satellites: Clues to Their Origins}.
\newblock {\em Science}, 318:\penalty0 1602--, December 2007.

\bibitem[{Quinn} et~al.(2010){Quinn}, {Perrine}, {Richardson}, and
  {Barnes}]{Quinn2010}
T.~{Quinn}, R.~P. {Perrine}, D.~C. {Richardson}, and R.~{Barnes}.
\newblock {A Symplectic Integrator for Hill's Equations}.
\newblock {\em \aj}, 139:\penalty0 803--807, February 2010.

\bibitem[{Rein} et~al.(2010){Rein}, {Lesur}, and
  {Leinhardt}]{ReinLesurLeinhardt2010}
H.~{Rein}, G.~{Lesur}, and Z.~M. {Leinhardt}.
\newblock {The validity of the super-particle approximation during planetesimal
  formation}.
\newblock {\em \aap}, 511:\penalty0 A69+, February 2010.

\bibitem[{Rein} and {Papaloizou}(2009)]{ReinPapaloizou2009}
H.~{Rein} and J.~C.~B. {Papaloizou}.
\newblock {On the evolution of mean motion resonances through stochastic
  forcing: fast and slow libration modes and the origin of HD 128311}.
\newblock {\em \aap}, 497:\penalty0 595--609, April 2009.

\bibitem[{Sei{\ss}} et~al.(2005){Sei{\ss}}, {Spahn}, {Srem{\v c}evi{\'c}}, and
  {Salo}]{Seiss2005}
M.~{Sei{\ss}}, F.~{Spahn}, M.~{Srem{\v c}evi{\'c}}, and H.~{Salo}.
\newblock {Structures induced by small moonlets in Saturn's rings: Implications
  for the Cassini Mission}.
\newblock {\em \grl}, 32:\penalty0 11205--+, June 2005.

\bibitem[{Spahn} and {Srem{\v c}evi{\'c}}(2000)]{Spahn2000}
F.~{Spahn} and M.~{Srem{\v c}evi{\'c}}.
\newblock {Density patterns induced by small moonlets in Saturn's rings?}
\newblock {\em \aap}, 358:\penalty0 368--372, June 2000.

\bibitem[{Srem{\v c}evi{\'c}} et~al.(2002){Srem{\v c}evi{\'c}}, {Spahn}, and
  {Duschl}]{Sremcevic2002}
M.~{Srem{\v c}evi{\'c}}, F.~{Spahn}, and W.~J. {Duschl}.
\newblock {Density structures in perturbed thin cold discs}.
\newblock {\em \mnras}, 337:\penalty0 1139--1152, December 2002.

\bibitem[{Tiscareno} et~al.(2008){Tiscareno}, {Burns}, {Hedman}, and
  {Porco}]{Tiscareno2008}
M.~S. {Tiscareno}, J.~A. {Burns}, M.~M. {Hedman}, and C.~C. {Porco}.
\newblock {The Population of Propellers in Saturn's A Ring}.
\newblock {\em \aj}, 135:\penalty0 1083--1091, March 2008.

\bibitem[{Tiscareno} et~al.(2006){Tiscareno}, {Burns}, {Hedman}, {Porco},
  {Weiss}, {Dones}, {Richardson}, and {Murray}]{Tiscareno2006}
M.~S. {Tiscareno}, J.~A. {Burns}, M.~M. {Hedman}, C.~C. {Porco}, J.~W. {Weiss},
  L.~{Dones}, D.~C. {Richardson}, and C.~D. {Murray}.
\newblock {100-metre-diameter moonlets in Saturn's A ring from observations of
  propeller structures}.
\newblock {\em \nat}, 440:\penalty0 648--650, March 2006.

\bibitem[{Tiscareno} et~al.(2010){Tiscareno}, {Burns}, {Sremcevic}, {Beurle},
  {Hedman}, {Cooper}, {Milano}, {Evans}, {Porco}, {Spitale}, and
  {Weiss}]{Tiscareno2010}
M.~S. {Tiscareno}, J.~A. {Burns}, M.~{Sremcevic}, K.~{Beurle}, M.~M. {Hedman},
  N.~J. {Cooper}, A.~J. {Milano}, M.~W. {Evans}, C.~C. {Porco}, J.~N.
  {Spitale}, and J.~W. {Weiss}.
\newblock {Directly Observing the Orbital Evolution of Disk-embedded Masses}.
\newblock In {\em Bulletin of the American Astronomical Society}, volume~41 of
  {\em Bulletin of the American Astronomical Society}, pages 939--+, May 2010.

\end{thebibliography}
